\documentclass[10pt,journal]{IEEEtran}

\ifCLASSOPTIONcompsoc
  \usepackage[nocompress]{cite}
\else
\usepackage{cite}
\fi

\usepackage{times}
\usepackage{epsfig}
\usepackage{graphicx}
\usepackage{amsmath}
\usepackage{amssymb}
\usepackage{multirow}
\usepackage{pifont}
\usepackage{booktabs}
\usepackage{booktabs,multirow}
\usepackage{graphics}
\usepackage{threeparttable}
\usepackage{color}
\usepackage[normalem]{ulem}
\usepackage{multirow}
\usepackage{float}
\usepackage{amsfonts}
\usepackage{bm}
\usepackage{bbm}
\usepackage{subfig}
\usepackage{enumitem}
\usepackage[numbers]{natbib}
 \usepackage{url}
\usepackage{array}
\usepackage[table]{xcolor}
\usepackage{colortbl}
\definecolor{bblue}{rgb}{0,150,230}
\definecolor{mygray}{gray}{.9}
\definecolor{myy}{RGB}{126,95,0}
\usepackage{bm}
\newcolumntype{I}{!{\vrule width 1pt}}

\usepackage{bbding}
\usepackage{pifont}
\usepackage{wasysym}
\usepackage{amssymb}
\definecolor{ggray}{RGB}{127,127,127}

\newcommand{\eg}[1]{\textit{e.g.,}}
\newcommand{\ie}[1]{\textit{i.e.,}}
\newcommand{\etc}[1]{\textit{etc}}

\makeatletter
\newcommand{\thickhline}{%
	\noalign {\ifnum 0=`}\fi \hrule height 1pt
	\futurelet \reserved@a \@xhline
}
\makeatother

\newcommand{\figref}[1]{Fig.\!~\ref{#1}}

\DeclareMathOperator*{\Motimes}{\text{\raisebox{0.25ex}{\scalebox{0.6}{$\bigotimes$}}}}

\usepackage[pagebackref=true,breaklinks=true,letterpaper=true,colorlinks,bookmarks=false]{hyperref}

\usepackage{hyperref}

\usepackage{cleveref}
\crefname{section}{}{§§}
\Crefname{section}{}{§§}

\usepackage{caption}
\begin{document}

\title{Exploring Separable Attention for Multi-Contrast MR Image Super-Resolution}
\author{Chun-Mei Feng, Yunlu Yan, Kai Yu, Yong Xu, \IEEEmembership{Senior Member, IEEE},  Jian Yang, \IEEEmembership{Senior Member, IEEE}, \\Ling Shao, \IEEEmembership{Fellow, IEEE} and Huazhu Fu, \IEEEmembership{Senior Member, IEEE}

\thanks{The work was supported by Shenzhen Science and Technology Innovation Committee (NO.~GJHZ20210705141812038), A*STAR Career Development Fund (IHPC/CI/G22-003), and AME Programmatic Fund (A20H4b0141).}

\thanks{C.-M.~Feng is with the Shenzhen Key Laboratory of Visual Object Detection and Recognition, Harbin Institute of Technology (Shenzhen), 518055, China, and also with the Institute of High Performance Computing, A*STAR, Singapore 138632.~(Email: strawberry.feng0304@gmail.com)}
\thanks{Y.~Yan, and Y.~Xu are with the Shenzhen Key Laboratory of Visual Object Detection and Recognition, Harbin Institute of Technology (Shenzhen), 518055, China.~(Email: yongxu@ymail.com).}
\thanks{J.~Yang is with the PCA Laboratory, Key Laboratory of Intelligent Perception and Systems for High-Dimensional Information of Ministry of Education, Nanjing University of Science and Technology, Nanjing 210094, China, and also with the Jiangsu Key Laboratory of Image and Video Understanding for Social Security, School of Computer Science and Engineering, Nanjing University of Science and Technology, Nanjing 210094, China (e-mail: csjyang@njust.edu.cn).}
\thanks{L.~Shao is with Terminus Group, China. (Email: ling.shao@ieee.org).}
\thanks{K.~Yu and H.~Fu is with the Institute of High Performance Computing, A*STAR, Singapore 138632. (E-mail: hzfu@ieee.org).}

\thanks{Corresponding author: \textit{Yong Xu and Huazhu Fu}.}

}

\markboth{IEEE Transactions on Neural Networks and learning Systems}%
{Shell \MakeLowercase{\textit{et al.}}: Bare Demo of IEEEtran.cls for Computer Society Journals}

\IEEEtitleabstractindextext{%
\begin{abstract}

Super-resolving the Magnetic Resonance (MR) image of a target contrast under the guidance of the corresponding auxiliary contrast, which provides additional anatomical information, is a new and effective solution for fast MR imaging. However, current multi-contrast super-resolution (SR) methods tend to concatenate different contrasts directly, ignoring their relationships in different clues, \eg, in the high-intensity and low-intensity regions. In this study, we propose a separable attention network (comprising high-intensity priority attention and low-intensity separation attention), named SANet. Our SANet could explore the areas of high-intensity and low-intensity regions in the ``forward'' and ``reverse'' directions with the help of the auxiliary contrast, while learning clearer anatomical structure and edge information for the SR of a target-contrast MR image. SANet provides three appealing benefits: (1) It is the first model to explore a separable attention mechanism that uses the auxiliary contrast to predict the high-intensity and low-intensity regions regions, diverting more attention to refining any uncertain details between these regions and correcting the fine areas in the reconstructed results. (2) A multi-stage integration module is proposed to learn the response of multi-contrast fusion at multiple stages, get the dependency between the fused representations, and boost their representation ability. (3) Extensive experiments with various state-of-the-art multi-contrast SR methods on fastMRI and clinical \textit{in vivo} datasets demonstrate the superiority of our model.

\end{abstract}

}

\maketitle
\IEEEdisplaynontitleabstractindextext
\IEEEpeerreviewmaketitle

\section{Introduction}

\label{sec:introduction}
Since Magnetic Resonance (MR) imaging has the advantages of being non-invasive, non-radiative and carried out \textit{in vivo}, it has become one of the most widely used imaging methods for disease diagnosis and treatment planning~\cite{wang2016accelerating,wang2020deepcomplexmri,feng2021specificity}. However, due to the physical nature of the MR data acquisition process, the scan time can take up to tens of minutes~\cite{feng2021DONet,feng2021dual}. Further, the quality of MR images obtained in clinical practice may be inadequate due to the patient's involuntary physiological movement (\eg, heart beating and breathing) during the acquisition process~\cite{feng2021task}. This is especially an issue when the scans are taken following multiple protocols that require a long echo time (TE) or repetition time (TR). These scans thus often yield poor diagnostic quality, since medical images must be examined at high resolution (HR) with rich structural details. If we could reconstruct HR images from low-resolution (LR) inputs, we would have the potential to achieve better spatial resolution, in shorter scan times~\cite{feng2022multi,feng2021multi}.

\begin{figure}[t]
\centering
  \includegraphics[width=0.50\textwidth]{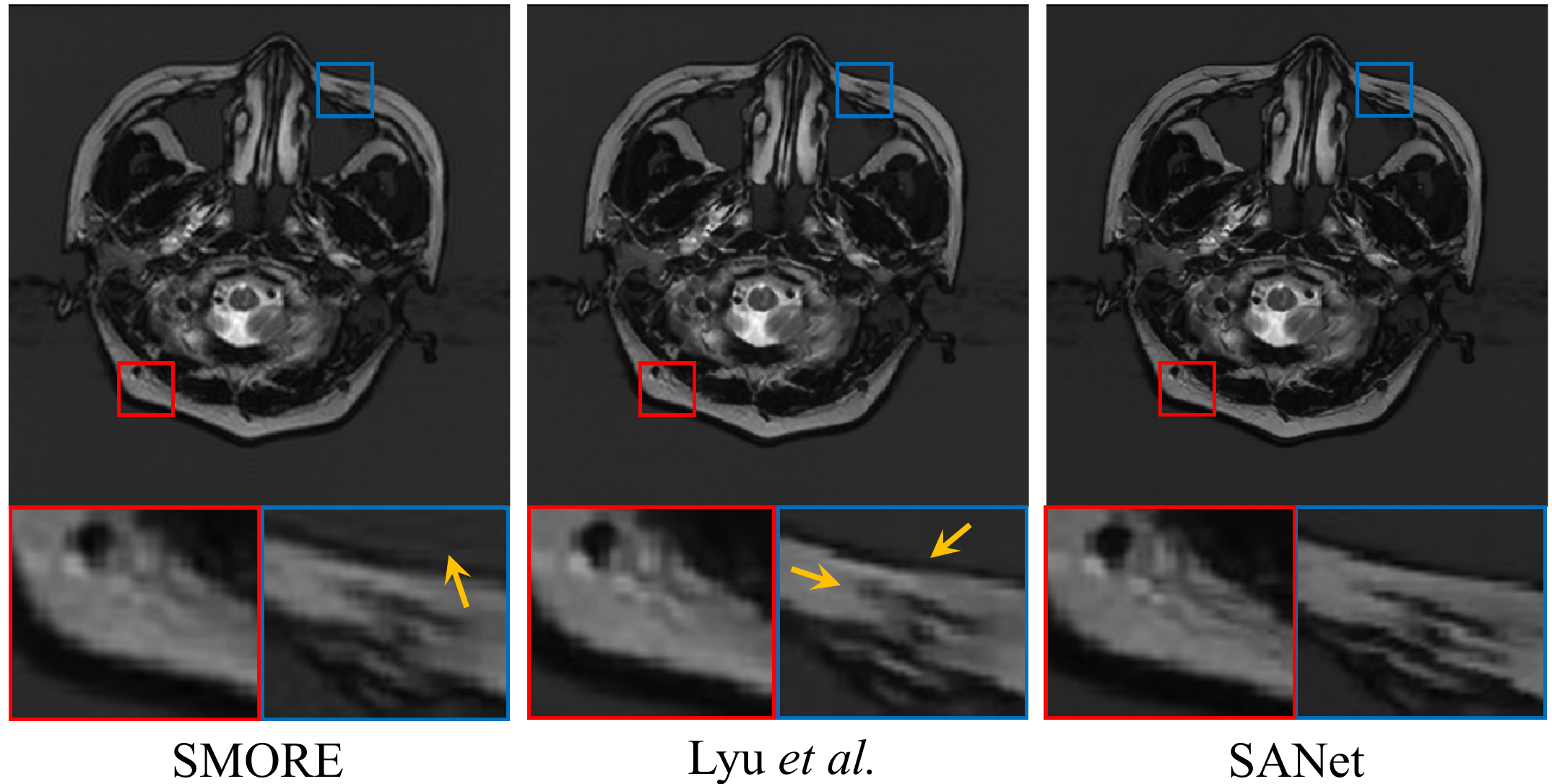}
  \caption{Comparison between single-contrast method SMORE~\cite{zhao2018deep}), multi-contrast method Lyu \textit{et al.}~\cite{zeng2018simultaneous}, and our SANet on \textit{in vivo} datasets. We effectively reconstruct the target contrast image by exploring the complementary information between the \textit{clear anatomical structure and boundary details} provided by the \textit{auxiliary contrast} from the separable attention. The visual comparison displays that our method can effectively recover the anatomical information. (\textbf{Zoom in for best view.})}
  \label{movitation}
\end{figure}

Super-resolution (SR) technology, which could enhance the image resolution without any hardware updates~\cite{zhang2020deep,zhang2018learning,zhang2019deep}, has been extensively studied to overcome the challenge of obtaining HR MR image scans~\cite{jiang2021fa}. However, the two most basic methods of SR interpolation~\cite{zhang2019deep,zhang2021plug,zhang2018ffdnet}, \ie, bicubic and b-spline interpolation, inevitably lead to blurred edges and blocking artifacts. To solve this problem, several methods have been proposed including traditional and deep learning based methods~\cite{shi2015lrtv,hardie2007fast,tourbier2015efficient,bhatia2014super,oktay2016multi,pham2017brain,mcdonagh2017context,chen2018brain,chaudhari2018super,lyu2018super}. However, most methods work on mono-contrast images to restore HR images, resulting in the multi-contrast information being ignored and bad image quality. (see the single-contrast method SMORE in~\figref{movitation}).

In clinical practice, different MR imaging settings will produce various multi-contrast images, \eg, T1 and T2 weighted images (T1WIs and T2WIs), as well as proton density and fat-suppressed proton density weighted images (PDWIs and FS-PDWIs), which can provide complementary information to each other~\cite{rousseau2010non,jafari2014mri}. For the same test case, these different-contrast images will have the same anatomical structure if they are aligned. However, they have different properties and acquisition times. For example, morphological and structural information can be provided by T1WIs, while edema and inflammation can be described by T2WIs. Further, PDWIs provide information on structures such as articular cartilage, and display high signal-to-noise ratios (SNRs) for tissues that have little difference in their PDWIs. Meanwhile, FS-PDWIs can inhibit fat signals and highlight the contrast between tissue structures, such as cartilage ligaments~\cite{chen2015accuracy}. During the acquisition process, T2WIs usually have a longer repetition time (TR) and echo time (TE) than T1WIs, while PDWIs usually have a shorter TR and TE than FS-PDWI. It should be noted that in the actual acquisition process, \textbf{\textit{multiple contrasts are often acquired together}} to facilitate the comprehensive diagnosis of disease. Thus, an intuitive idea is to use the contrast with the shorter acquisition time as additional information to assist the contrast with the slower imaging speed, the overall acquisition time and difficulty will be greatly reduced~\cite{zeng2018simultaneous,lyu2020multi}. For instance, additional information from T1WIs or PDWIs can be used as auxiliary contrasts to help the generation of target contrasts, \eg, T2WIs or FS-PDWIs~\cite{xiang2018deep,zheng2017multi,lu2015mr}.

Assisted acquisition of difficult contrasts through easily accessible contrasts has been proven to reduce clinical acquisition time and difficulty~\cite{zheng2017multi,zheng2018multi,zeng2018simultaneous,lyu2020multi}. They aim to restore the target contrast which with lower resolution through the auxiliary one while with full resolution. For example, Zheng \textit{et al.}~estimate the statistical information to restore a target SR image with the help of another contrast image~\cite{zheng2017multi,zheng2018multi}. In work~\cite{zeng2018simultaneous}, a deep convolutional neural network (CNN) is utilized to  produce the multi-contrast SR images, where the one contrast applies an HR image to assist another one~\cite{zeng2018simultaneous}. A progressive network is introduced in~\cite{lyu2020multi}, which is combined with the HR auxiliary contrast images for the target contrast SR.

However, though significant progress has been achieved in multi-contrast MR image SR, \textit{existing methods still have the following problems}: \textbf{(1)} Current methods typically restore the entire HR image, but the low-intensity area is often filled with zeros or contains noise that is meaningless for clinical diagnosis. Nevertheless, exploring the relationship between the high-intensity and low-intensity regions can improve the performance, because inaccurate reconstruction results are usually produced on the structural edges between the two regions. This has never been considered in previous studies (see the area indicated by the yellow arrow in \figref{movitation}). \figref{movitation} (a) and (b) show the results of the multi-modal MR image SR methods, but the recovery of anatomical boundary details of these methods is significantly lower than our method (\figref{movitation} (c)). Therefore, exploring the relationship between the high-intensity and low-intensity regions can improve the performance. \textbf{(2)} Current multi-contrast approaches ignore the interaction between different modalities at various stages. However, during the target-contrast restoration, if the feature maps of different layers are treated equally, some detailed anatomical textures will be lost in the reconstructed image. Therefore, we speculate that MR image SR is still possible to achieve significant improvement by considering both \textbf{(1)} and \textbf{(2)}.

With this insight, we take a further step towards exploring the \textbf{boundary details} between \textit{high-intensity anatomical structure and low-intensity regions} of an MR image separately for SR from the ``forward'' and ``reverse'' perspectives. The terms ``forward'' and ``backward'' denote refinement from the object to the high-intensity region and from the low-intensity region to the object, respectively. Thereby, the network can concentrate on the uncertain details between the two regions thanks to the exploration of two opposing directions. Specifically, we propose a separable attention network to fuse MR images of different contrasts and generate stronger target features. Our motivation stems from the fact that during the SR process, the anatomical contours are typically learned first, while the boundary details of some structures are more difficult to learn~\cite{van2012super}. These detailed areas are usually important for the diagnosis of lesions~\cite{almeida2020deep,mishra2018ultrasound}. Therefore, the \textit{recovery of anatomical boundary details} (see our SANet in~\figref{movitation}) is the main goal of MR image SR. With this mind, the primary focus of our work is to explore the high-intensity and low-intensity regions knowledge of the auxiliary contrast MR image separately, so as to \textit{shift more attention towards refining the uncertain details between the two regions}. This has never yet been explored within our knowledge. In addition, the features of each stage are interacted and weighted to yield a comprehensive feature, which can guide the learning of the target features.
Overall, our main \textbf{contributions} are three-fold: 
\begin{itemize}
\item {A novel multi-contrast fusion network, named SANet, is designed for multi-contrast MR image SR. It uses separable attention to assist the target contrast in recovering clear anatomical information (\eg~bones, tissues, blood vessels, and muscles) by exploring the \textit{uncertain details} between the high-intensity and low-intensity regions provided by the \textit{auxiliary contrast}.} 
\item {A multi-stage integration module is developed to explore the \textit{response of the multi-contrast fusion at different stages}. This allows us to mine the dependency relationship between the fused features and boost their representation ability.}
\item {Extensive experiments on fastMRI and two clinical datasets demonstrate the significant performance improvements of SANet against current state-of-the-art methods.\footnote{The code is released at \textcolor{red}{\url{https://github.com/chunmeifeng/SANet}}.}}
\end{itemize}

Compared with the original conference version~\cite{feng2021MINet}, in this paper, we provide several important extensions. (1) We generalize the architecture to a new version (\ie, SANet). This is achieved by our new separable attention compensation module, which aims to help the target contrast restore clear anatomical information by exploring the uncertain details between high-intensity and low-intensity regions provided by the auxiliary contrast (\S\ref{sec:separable}). (2) We also propose a high-intensity priority attention to gradually refine the anatomical structure in a ``forward'' manner, and a low-intensity separation attention to strengthen the edge information of the two regions in a ``reverse'' manner. (3) We provide more insightful experiments to evaluate the mechanism of our separable attention module in Sections \S\ref{sec:separable} and \S\ref{sec:ex}.

\section{Related Work}\label{sec:related_work}
\subsection{Deep Learning for MR Image SR}
SR techniques are widely used in MR imaging as a form of post-processing to enhance image quality without changing the acquisition hardware. However, traditional methods, such as low rank~\cite{shi2015lrtv}, iterative deblurring algorithms~\cite{hardie2007fast,tourbier2015efficient}, and dictionary learning methods~\cite{bhatia2014super,yue2019robust}, provide very limited information within a single image. To achieve better restoration, some techniques use additional training datasets to capture the structural information between training images~\cite{rueda2013single,wang2014sparse,zhang2015mr,jia2015single}. Further, as CNNs delve deeper into computer vision tasks, many SR techniques employing deep learning for MR imaging have been developed~\cite{litjens2017survey,yang2018admm}. For example, Oktay \textit{et al.}~computed the mapping between LR and HR cardiac MR images using CNNs to restore heart volumes~\cite{oktay2016multi}. Pham \textit{et al.}~also applied SR technology based on deep learning, using patches of other HR scans to enhance fetal brain MR images~\cite{pham2017brain}. McDonagh \textit{et al.}~used a residual CNN architecture with context-sensitive upsampling to generate an improved image with sharp edges and anatomical details~\cite{mcdonagh2017context}. 3D residual networks have also been used to generate thin-slice MR images~\cite{chaudhari2018super}. 

To restore an enhanced image with sharp details from a single degraded one, Chen \textit{et al.}~applied a densely connected SR network~\cite{chen2018brain}. More recently, generative adversarial network (GAN) frameworks~\cite{zhang2019self} used for CT denoising have been transferred to MR image SR~\cite{lyu2018super}. For instance, a fast and accurate GAN-based network was proposed in~\cite{chen2020mri} to explore the SR of multi-level dense connections. Subsequently, more powerful discriminators, such as the pyramid pooling discriminator, have been proposed to recover details at different scales for the SR of MR images~\cite{wang2020enhanced}. However, the above methods focus on a single contrast, such as a T1WI or T2WI, when restoring HR images, ignoring the complementary multi-contrast information of the MR image data.

\subsection{Multi-Contrast MR Image Representation}

In clinical practice, MR images are usually acquired with multiple contrasts for comprehensive diagnosis. Each contrast reflects a different type of tissue, but with the same anatomical structures. Benefiting from this, multi-contrast methods have been proposed to learn powerful representations for various MR image tasks, such as segmentation and SR. For example, Huo \textit{et al.}~used T1WI and T2WI scans together to improve the performance of MR image segmentation~\cite{huo2018splenomegaly}. Since multi-contrast images provide richer information than single-contrast ones, various MR image segmentation tasks, such as CNN-based and GAN-based methods, are built on these~\cite{despotovic2010brain,valindria2018multi,mondal2018few}. Different from the multi-contrast MR image segmentation task, in the SR task the images are divided into auxiliary and target contrasts. The auxiliary contrast, which has a shorter acquisition time, is used to guide the restoration of the target contrast. For example, Rousseau \textit{et al.}~used the anatomical priors from a paired auxiliary image to super-resolve the target image~\cite{rousseau2010non}. Manjón~\textit{et al.}~restored the LR T2WI with the help of the auxiliary T1WI of the same subject~\cite{manjon2010mri}. Meanwhile, similar anatomical structures from the auxiliary contrast can be used to obtain an SR image from its corresponding LR image~\cite{zheng2017multi,zheng2018multi,lu2015mr,jafari2014mri}. Recently, CNN-based methods, \eg, UNet and GAN-based methods, have been used to super-resolved the target contrast image from the help of auxiliary one \cite{zeng2018simultaneous,lyu2020multi}. However, previous methods simply use the auxiliary contrast as the prior data, without exploring the potential correlation between the different contrasts. In addition, the LR MR image scans in these works are simulated by downsampling the magnitude images, which violates the imaging principle of MR images, leaving the effectiveness to be verified. Besides, current various computer vision works based on attention mechanisms have obtained outstanding results~\cite{shu2022expansion,tang2019coherence,shu2021spatiotemporal}. For example, they aggregate multi-modal features in modal and channel-wise ways~\cite{shu2022expansion}, quantify the contribution of a specific motion by measuring the consistency between the human motion and the overall activity~\cite{tang2019coherence}, and capture spatial coherence among joints using a skeleton-joint co-attention mechanism~\cite{shu2021spatiotemporal}. This motivates us to enhance the aggregation of both modalities in the same way.

In this work, we build our model by simulating an actual LR image to demonstrate the appealing properties of SANet for the SR of multi-contrast MR images. Based on this, we propose a separable attention network that learns the uncertainty between the high-intensity and low-intensity regions of the auxiliary contrast to improve the restoration performance of the target contrast. 

\section{Methodology}

\subsection{Architecture Overview}\label{arc}

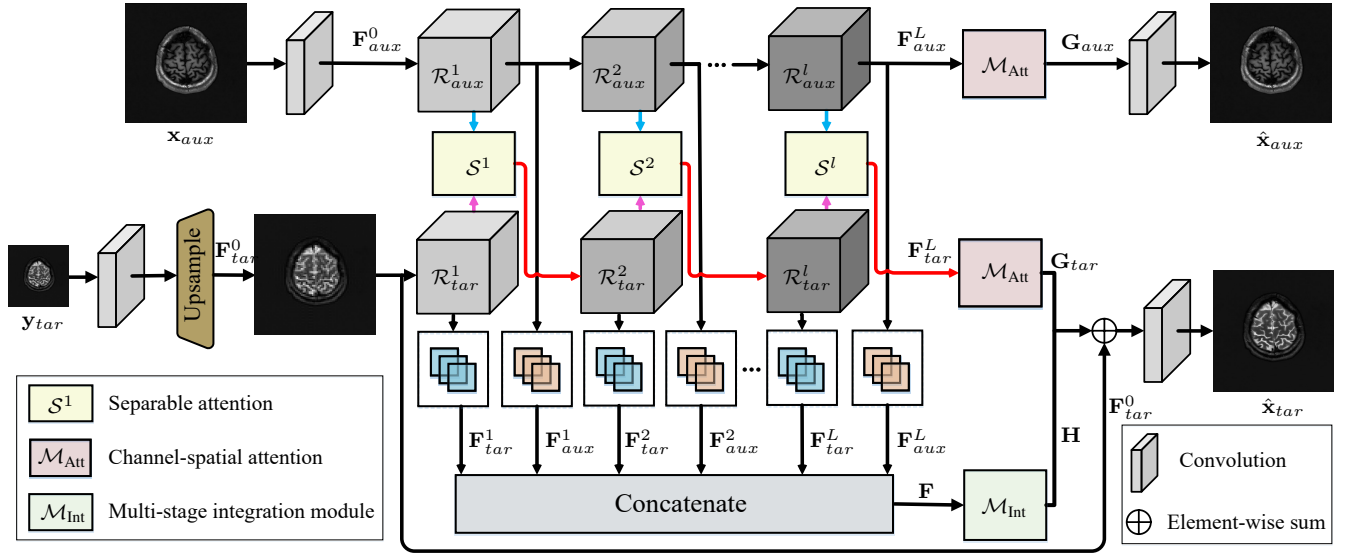
\begin{figure*}[t]
\centering
  \includegraphics[width=0.98\textwidth]{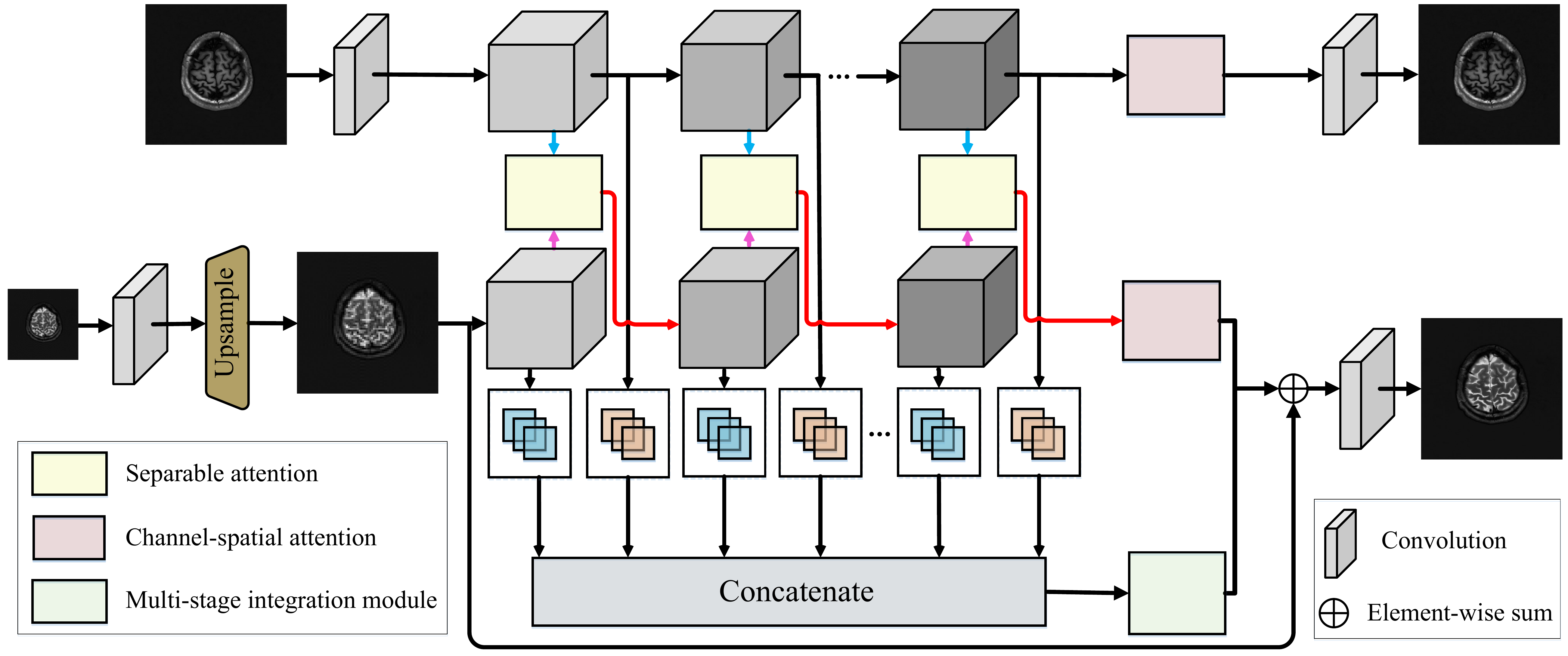}
  \put(-443,158){\footnotesize $\mathbf{x}_{aux}$}
  \put(-498,88){\footnotesize $\mathbf{y}_{tar}$}
  \put(-31,156){\footnotesize $\hat{{\mathbf x}}_{aux}$}
  \put(-29,55){\footnotesize $\hat{{\mathbf x}}_{tar}$} 
  \put(-343,180){\footnotesize $\mathcal{R}_{aux}^1$} 
  \put(-282,180){\footnotesize $\mathcal{R}_{aux}^2$}
  \put(-210,180){\footnotesize $\mathcal{R}_{aux}^l$}
  \put(-343,104){\footnotesize $\mathcal{R}_{tar}^1$}
  \put(-280,104){\footnotesize $\mathcal{R}_{tar}^2$}
  \put(-210,104){\footnotesize $\mathcal{R}_{tar}^l$}
  \put(-330,145){\footnotesize $\mathcal{S}^1$}
  \put(-268,145){\footnotesize $\mathcal{S}^2$}
  \put(-198,145){\footnotesize $\mathcal{S}^l$}  
  \put(-135,106){\footnotesize $\mathcal{M}_{\text{Att}}$}
  \put(-135,184){\footnotesize $\mathcal{M}_{\text{Att}}$}
  \put(-135,17){\footnotesize $\mathcal{M}_{\text{Int}}$}
  \put(-105,193){\footnotesize $\mathbf{G}_{aux}$}
  \put(-109,110){\footnotesize $\mathbf{G}_{tar}$}
  \put(-105,44){\footnotesize $\mathbf{H}$}
  \put(-159,23){\footnotesize $\mathbf{F}$}
  \put(-425,114){\footnotesize $\mathbf{F}_{tar}^{0}$}
  \put(-329,42){\footnotesize $\mathbf{F}_{tar}^1$}
  \put(-300,42){\footnotesize $\mathbf{F}_{aux}^1$}
  \put(-270,42){\footnotesize $\mathbf{F}_{tar}^2$}  
  \put(-238,42){\footnotesize $\mathbf{F}_{aux}^2$}  
  \put(-200,42){\footnotesize $\mathbf{F}_{tar}^L$}  
  \put(-167,42){\footnotesize $\mathbf{F}_{aux}^L$}  
  \put(-373,194){\footnotesize $\mathbf{F}_{aux}^0$}
  \put(-87,55){\footnotesize $\mathbf{F}_{tar}^0$}
  \put(-163,114){\footnotesize $\mathbf{F}_{tar}^L$}
  \put(-167,194){\footnotesize $\mathbf{F}_{aux}^L$}
  \put(-488,55){\footnotesize $\mathcal{S}^1$}
  \put(-492,36){\footnotesize $\mathcal{M}_{\text{Att}}$}
  \put(-491,15){\footnotesize $\mathcal{M}_{\text{Int}}$}
  \caption{\textbf{Illustration of our SANet framework}. Our model is made up of two branches: a target contrast branch and an auxiliary contrast branch. We input the features $\mathbf{x}_{aux}$ and $\mathbf{x}_{tar}$ of two contrasts to the residual groups $\mathcal{R}_{aux}^l$ and $\mathcal{R}_{tar}^l$ to learn a multi-stage feature representation. In each stage, we aggregate the features of the two contrasts with separable attention $\mathcal{S}^l$. Then we use the channel-spatial attention module $\mathcal{M}_{\text{Att}}$ to enhance them, and a multi-stage integration module $\mathcal{M}_{\text{Int}}$ to learn the correlation among different stages.}
  \label{fig21} 
\end{figure*}

Let $\mathbf{x}\in\mathbb{R}^{H\times W}$ represent a fully sampled HR MR image. We can obtain the corresponding LR image $\mathbf{y}\in\mathbb{R}^{\frac{H}{s}\times \frac{W}{s}}$ ($s$ is the scale factor) by truncating the outer part of the fully sampled $k$-space according to the desired scale factor and then applying a 2D fast Fourier transform (FFT) $\mathcal{F}$~\cite{chen2018efficient}. Thus, the down-scaled LR image can be obtained. This process mimics the real image acquisition process, making it different from current SR methods for MR images~\cite{chen2018efficient}, which simply add Gaussian blurs or noise to the downsampled amplitude image to obtain the LR one~\cite{pham2017brain}. Therefore, this approach can accurately evaluate the effectiveness of the SR models. The SR process for MR images aims to learn a CNNs that can restore an SR image $\hat{{\mathbf x}}$ that is similar to its corresponding HR image $\mathbf{x}$. Current efforts aim to solve this problem by directly super-resolve $\hat{\mathbf{x}}$ from $\mathbf{x}$, followed by the existing multi-contrast works~\cite{zheng2017multi,zheng2018multi,zeng2018simultaneous,lyu2020multi}, we try to solve it from a multi-contrast perspective by adding an additional contrast image with the same anatomical structure, which yields higher quality details.

In~\figref{fig21}, we employ a two-branch structure with an HR auxiliary contrast to enhance the qualitiy of the LR target contrast. Both the HR auxiliary contrast $\mathbf{x}_{aux}$ and LR target contrast $\mathbf{y}_{tar}$ are accepted as the input of our SANet. We send the two contrasts to two different branches to learn a deep combination. To mine the complementary information of the two different contrast, we explore the high-intensity and low-intensity cues in both directions under the guidance of the auxiliary contrast, so that the target branch can learn the edges and fine structure of the image from the auxiliary contrast. Further, to explore the dependencies of multi-contrast features in different stages, we propose a multi-stage integration module that aims to yield a more powerful holistic feature representation for SR.

\subsubsection{Feature Extraction}
We employ the target and auxiliary contrast information in two streams. Specifically, we use a $3\times3$ convolutional layer to the auxiliary stream to yield an initial feature $\mathbf{F}_{aux}^0$. Similarly, the target stream is processed by a convolutional layer and a sub-pixel convolution~\cite{wang2020deep} $\operatorname{U}_{\uparrow}$ to obtain the enlarged feature $\mathbf{F}_{tar}^0$:
\begin{align}
    \mathbf{F}_{aux}^0&=\operatorname{Conv}\left(\mathbf{x}_{aux}\right),\\
    \mathbf{F}_{tar}^0&=\operatorname{U}_{\uparrow}\left({\operatorname{Conv}\left(\mathbf{y}_{tar}\right)}\right).
\end{align}
We use this pre-upsampling strategy because it provides two advantages~\cite{wang2020deep}: (1) The CNNs only need to refine the coarse HR image, which reduces the difficulty of learning. (2) The features in both streams being of identical size allows the target contrast to inherit the clear structure and edge information from the bidirectional high-intensity and low-intensity cues in the auxiliary contrast.

\subsubsection{Multi-Stage Feature Representation}
Unlike previous works~\cite{zheng2017multi,zheng2018multi,zeng2018simultaneous,lyu2020multi}, we explore the cross-contrast fusion of auxiliary and target features at multiple stages, rather than fusing them in an early stage or the input. Specifically, the multi-stage feature representations are learned through a cascade of residual groups $\mathcal{R}$ ~\cite{zhang2018image}. In particular, in the $l^{th}$ stage, we obtain the intermediate features as follows:
\begin{align}
    \mathbf{F}_{aux}^{l}&=\mathcal{R}_{aux}^l(\mathbf{F}_{aux}^{l}), l \in\{1, \ldots, L\}\\
    \mathbf{F}_{tar}^{l}&=\mathcal{R}_{tar}^l(\mathcal{S}^l(\mathbf{F}_{aux}^{l},\mathbf{F}_{tar}^{l})).\label{eq:2}
\end{align}
Here, $\mathcal{R}_{aux}^l$ and $\mathcal{R}_{tar}^l$ represent the residual groups in the $l^{th}$ stage of each respective branch. $\mathcal{S}^l$ represents the separable attention module in the $l^{th}$ stage, which will be introduced in \S\ref{sec:separable}. Through Eq.~\eqref{eq:2}, our model uses the clear structural features of the auxiliary contrast to bidirectionally explore the high-intensity and low-intensity cues. Instead of the current method which uses only the representation of the final stage for SR~\cite{lyu2020multi,lim2017enhanced}, we try to store the features of multiple stages: 
\begin{equation}\label{eq:3}
    \mathbf{F} = [\mathbf{F}_{aux}^1, \mathbf{F}_{tar}^1, \mathbf{F}_{aux}^2, \mathbf{F}_{tar}^2, \cdots \mathbf{F}_{aux}^L, \mathbf{F}_{tar}^L].
\end{equation}
Here, $\mathbf{F}\!\in\mathbb{R}^{2L\times\!H\times\!W\times\!C}$ is a multi-stage feature representation obtained by concatenating all the intermediate features, and $L$ represents the number of residual groups, $C$ is the number of channels.

\subsubsection{Multi-Contrast Feature Enhancement}
Eq.~\eqref{eq:3} aggregates the fusion features of multiple stages and has a more comprehensive representation. However, each feature is independent, and the relationship between them has not been mined. Thus, we design a multi-stage integration module $\mathcal{M}_{\text{Int}}$ (\S\ref{sec:ml}), which learns the relationships between the features of each stage and uses them to modulate each stage:
\begin{equation}
\mathbf{H} = \mathcal{M}_{\text{Int}}(\mathbf{F}),
\end{equation}
where $\mathbf{H}$ denotes the enriched representation integrated from various stages. To further enhance the feature representation, we design a channel-spatial attention module $\mathcal{M}_{\text{Att}}$ to modulate the feature $\mathbf{F}_{tar}^L$ as $\mathbf{G}_{tar}\!=\!\mathcal{M}_{\text{Att}}(\mathbf{F}_{tar}^L)$. Based on the spatial~\cite{woo2018cbam} and channel attention mechanisms~\cite{zhang2018image}, the $\mathcal{M}_{\text{Att}}$ reveals the responses from all dimensions of the feature maps. As can be seen from~\figref{fig23} (b), we send the 
$\mathbf{F}_{tar}^L$ to a 3D convolutional layer. To this end, an attention map $\mathbf{A}_{tt}$ is generated by capturing joint channels and spatial features. Formally, we have
\begin{equation}\label{eq:10}
\mathbf{G}_{tar}= \sigma\left(\mathbf A_{tt}\right) \Motimes \mathbf{F}_{tar}^L+\mathbf{F}_{tar}^L,
\end{equation}
where $\sigma$ represents the sigmoid operation, and $\Motimes$ is element-wise multiplication. Thus, Eq.~\eqref{eq:10} mechanism makes $\mathbf {F}_{tar}^L$ enhanced by the context in the feature.
We also use a residual layer to preserve information from the original feature. Similarly, the enhanced auxiliary contrast features $\mathbf{G}_{aux}$ can also be calculated using Eq.~\eqref{eq:10}.

\begin{figure}[t]
\centering
  \includegraphics[width=0.5\textwidth]{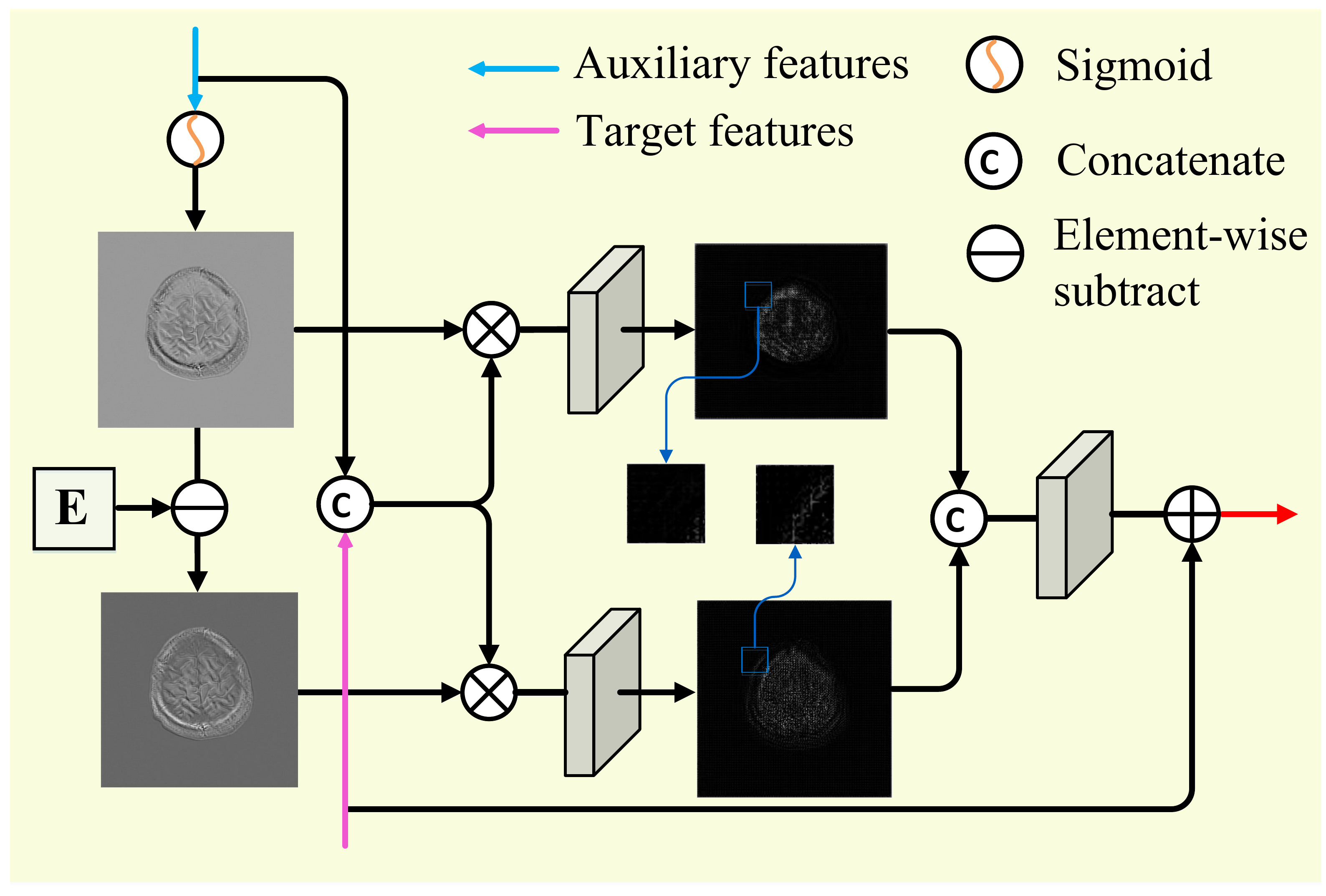}
  \put(-245,157){\footnotesize $\mathbf{F}_{aux}^l$}
  \put(-213,10){\footnotesize $\mathbf{F}_{tar}^l$}
  \put(-182,80){\footnotesize $\hat{\mathbf{F}}^{l}$}
  \put(-255,105){\footnotesize $\mathbf{A}_{H}^{l}$}
  \put(-255,35){\footnotesize $\mathbf{A}_{L}^{l}$}
  \put(-21,58){\footnotesize $\mathbf{R}^{l}$}
  
  \caption{Illustration of the proposed separable attention compensation module. $\mathbf{E}$ is a matrix in which all the elements are 1. $\mathbf{A}_{H}^{l}$ and $\mathbf{A}_{L}^{l}$ are the high-intensity priority attention (HP) map and the low-intensity attention (LS) map, respectively. $\mathbf{F}_{aux}^l$ and $\mathbf{F}_{tar}^l$ are the intermediate features of the two contrasts, $\hat{\mathbf{F}}^{l}$ is the fused feature of the $l^{th}$ stage, and $\mathbf{R}^{l}$ is the output of the separable attention module. The features obtained from the HP provide more anatomical structure (zoom in for best view), while the features obtained from the LS have more edge details (see the {\color{cyan}blue} box). In this way, our method can achieve significantly improvement in anatomical boundaries of the soft tissue (see the yellow arrows in~\figref{fig01} for more details).}
  \label{fig22} 
\end{figure}

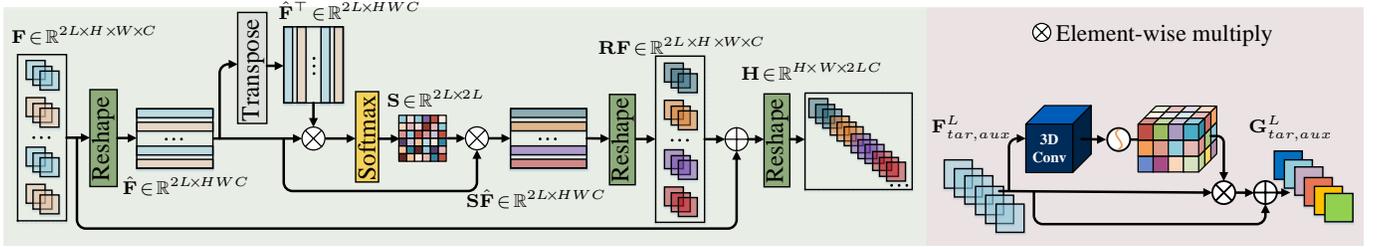
\begin{figure*}[!t]
\centering
  \includegraphics[width=1\textwidth]{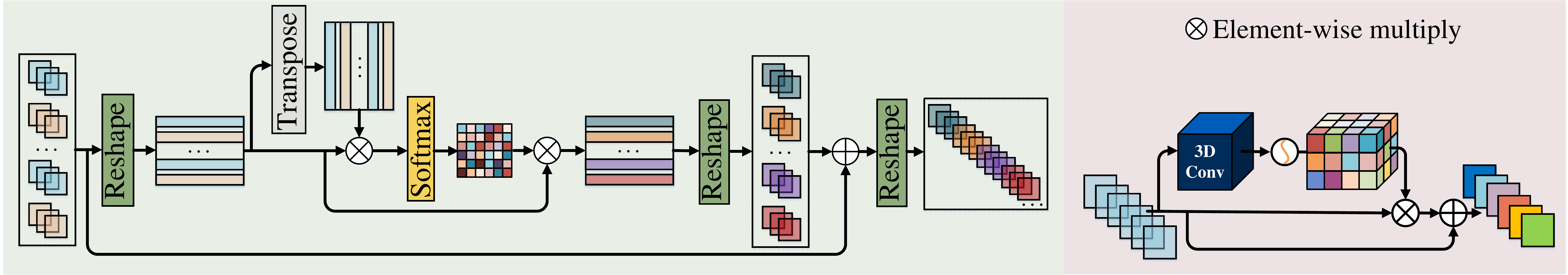}
  \put(-512,77){\scriptsize $\mathbf{F}\!\in\!\mathbb{R}^{2L\!\times\!H\times\!W\!\times\!C}$}
  \put(-470,20){\scriptsize $\hat{\mathbf{F}}\!\in\!\mathbb{R}^{2L\!\times\! HWC}$}
  \put(-411,86){\scriptsize $\hat{\mathbf{F}}^\top\!\in\!\mathbb{R}^{2L\!\times\! HWC}$}
  \put(-370,53){\scriptsize $\mathbf{S}\!\in\!\mathbb{R}^{2L\!\times\! 2L}$}
  \put(-340,15){\scriptsize $\mathbf{S}\hat{\mathbf{F}}\!\in\!\mathbb{R}^{2L\!\times\!HWC}$}
  \put(-236,63){\scriptsize $\mathbf{H}\!\in\!\mathbb{R}^{H\!\times W\!\times\!2LC}$}
  \put(-290,73){\scriptsize $\mathbf{R}{\mathbf{F}}\!\in\!\mathbb{R}^{2L\times \!H\times \!W\!\times\!C}$}   
  \put(-164,43){\scriptsize $\mathbf{F}_{tar,aux}^L$}  
  \put(-44,43){\scriptsize $\mathbf{G}_{tar,aux}^L$}  
  
  \caption{(a) Illustration of the proposed multi-stage integration module, where $\mathbf{F}$ is the multi-stage feature representation, $\mathbf{S}$ is the affinity matrix, and $\mathbf{H}$ is the comprehensive and holistic output feature representation. (b) Illustration of the proposed channel-spatial attention module, where  $\mathbf{F}_{tar,aux}^L$ is the input feature output by the residual group, and $\mathbf{G}_{tar,aux}^L$ are the output feature after a 3D convolutional layer.} 
  \label{fig23} 
\end{figure*}

\subsubsection{Super-Resolved Image}
Because our method is a form of pre-upsampling SR, a convolutional layer $\operatorname{Conv}_{tar}$ is used to convolve $\mathbf{G}_{tar}$ and $\mathbf H$ with residual compensation to obtain the final restored image $\hat{{\mathbf x}}_{tar}$:
\begin{equation}\label{eq:6}
\begin{aligned}
\hat{{\mathbf x}}_{tar}&=\operatorname{Conv}_{tar}(\mathbf{F}_{tar}^{0}\oplus\mathbf{G}_{tar}\oplus\mathbf{H}),\\
\hat{{\mathbf x}}_{aux}&=\operatorname{Conv}_{aux}({\mathbf{G}_{aux}}),
\end{aligned}
\end{equation}
where $\oplus$ is the element-wise summation. Besides, as can be seen from~\figref{fig21}, we also restore the auxiliary contrast $\hat{{\mathbf x}}_{aux}$ using a convolutional layer with input $\mathbf{G}_{aux}$. Note that this reconstruction procedure of the auxiliary contrast is only used to maintain consistency with the input.

\subsubsection{Loss Function}
Here, the $L_1$ loss is employed to evaluate the results of the target and auxiliary contrast. The overall loss function can be expressed as
\begin{equation}
L=\frac{1}{N} \sum_{n=1}^{N} \alpha\left\|\hat{{\mathbf x}}_{tar}^{n}-{{\mathbf x}}_{tar}^{n}\right\|_{1}+(1-\alpha)\left\|\hat{{\mathbf x}}_{aux}^{n}-{{\mathbf x}}_{aux}^{n}\right\|_{1},
\end{equation}
where $\alpha$ weights the trade-off between the reconstruction of the two contrasts, and $N$ is the number of training samples.

\subsection{Separable Attention Compensation}\label{sec:separable}
Since simple fusion of HR auxiliary contrast to assist the SR of target contrast cannot meet current requirements, we further explore the relationship between multi-contrast here~\cite{zheng2017multi,zheng2018multi,zeng2018simultaneous,lyu2020multi}. Given an HR auxiliary contrast, in this paper we aim to refine and reconstruct the target contrast using the \textit{clear structural information from the auxiliary contrast}. The distributions of the high-intensity and low-intensity regions in MR images are quite different. For example, the low-intensity region is usually all zero or contains some noise data, while the high-intensity region contains all the structural and diagnostic information. We design a separable attention module to learn the high-intensity and low-intensity features from two directions at each stage, and then jointly refine the uncertain details between the two regions. As can be seen in~\figref{fig22}, we compute the high-intensity priority attention (HP) map $\mathbf{A}_{H}^{l}$ using a sigmoid operation, while the low-intensity separation attention (LS) map $\mathbf{A}_{L}^{l}$ is generated by subtracting the HP attention map from matrix $\mathbf{E}$:

\begin{equation}
\begin{aligned}
&\mathbf{A}_{H}^{l}=\sigma\left(\mathbf{F}_{a u x}^{l}\right), \\
&\mathbf{A}_{L}^{l}=\mathbf{E}-\sigma\left(\mathbf{F}_{a u x}^{l}\right),
\end{aligned}
\end{equation}
where $\mathbf{E}$ is a matrix in which all the elements are 1. Then, we obtain the fused feature $\hat{\mathbf{F}}^{l}$ as follows:
\begin{equation}
\hat{\mathbf{F}}^{l}=\operatorname{Conv_{Re}}[\mathbf{F}_{aux}^l,\mathbf{F}_{tar}^l],
\end{equation}
where $\operatorname{Conv_{Re}}$ is a 1$\times$1 convolution with 32 output channels to reduce the computational cost. Then, we use the HP and LS, as well as the residual component, to weight the separate features in the two streams, which can be expressed as

\begin{equation}
\mathcal{R}^{l}=\mathcal{Q}_{R}\left(\left[\mathcal{Q}\left(\hat{\mathbf{F}}^{l} \Motimes \mathbf{A}_{H}^{l}\right), \mathcal{Q}\left(\hat{\mathbf{F}}^{l} \Motimes \mathbf{A}_{L}^{l}\right)\right]\right)+\mathbf{F}_{tar}^{l},
\end{equation}
where $[,]$ denotes the concatenation operation, and $\mathcal{Q}$ represents the 3$\times$3 convolutional layer and a ReLU layer for extracting features. Finally, we use a prediction layer $\mathcal{Q}_{R}$ with the same operation as $\mathcal{Q}$ to output a single-channel residual map. It is obvious that the features obtained from the HP provide more anatomical structure, while the features obtained from the LS have more edge details (see the blue box in~\figref{fig22}).

\subsection{Multi-Stage Integration Module}\label{sec:ml}
Although our separable attention aggregates the features in the intermediate stages, it would be interesting if we can explore the correlation among different stages. If we regard the different contrasts as different categories, then features of different stages can be considered as as responses to specific categories. To this end, we propose a multi-stage integration module $\mathcal{M}_{\text{Int}}$. As can be seen from~\figref{fig23} (a), the network can assign different weights to the features of different contrasts and stages according to the dependence between the auxiliary contrast features of each stage and other stages. As a result, the network not only focuses on the more informative stages of the target branch, but also emphasizes the stages that guide the SR of the target contrast.

The $\mathbf{F}\!\in\mathbb{R}^{2L\times\!H\times\!W\times\!C}$ in Eq.~\eqref{eq:3} has concatenates all intermediate representations together. To enable our multi-stage integration module to understand the correlation between each pair of multi-contrast features. First, we flatten $\mathbf{F}$ into a matrix representation $\hat{\mathbf{F}}\in\mathbb{R}^{2L\times HWC}$ for computational convenience. Then, we compute the correspondence between each pair of features $\hat{\mathbf{F}}_i\in\mathbb{R}^{HWC}$ and $\hat{\mathbf{F}}_j\in\mathbb{R}^{HWC}$ using the following bilinear model:
\begin{equation}
\begin{aligned}\small
\mathbf{S}&=\texttt{softmax}(\hat{\mathbf{F}}  \hat{\mathbf{F}}^\top) \\
&= \texttt{softmax}([\hat{\mathbf{{F}}}_1, \cdots, \hat{\mathbf{F}}_{2L}][\hat{\mathbf{F}}_1, \cdots, \hat{\mathbf{F}}_{2L}]^\top) \in [0,1]^{2L\times 2L}.
\end{aligned}
\end{equation}
Here, $\mathbf{S}$ represents the affinity matrix to store the similarity scores corresponding to each pair of features in $\hat{\mathbf{F}}$, \ie, the $(i,j)^{th}$ element of $\hat{\mathbf{F}}$ gives the similarity between $\hat{\mathbf{F}}_i$ and $\hat{\mathbf{F}}_j$. We use $\texttt{softmax}(\cdot)$ to normalize each row of the input. Then, attention summaries are computed as $\mathbf{S}\hat{\mathbf{F}}\in\mathbb{R}^{2L\times HWC}$ and used the following form to generate an enhanced representation: 
\begin{equation}
    \mathbf{H} = \texttt{reshape}(\mathbf{R}{\mathbf{F}} + \mathbf{F}) \in \mathbb{R}^{H\times W\times 2LC},
\end{equation}
where $\mathbf{R}{\mathbf{F}}$ is reshaped by $\mathbf{S}\hat{\mathbf{F}}$. 
In this way, our model is able to learn a more comprehensive and holistic feature representation $\mathbf{H}$ by exploring the relations between features of multi-contrast images at multiple stages.
Thus, with the help of multi-contrast related features at multiple stages, our model can capture a more holistic and comprehensive feature representation $mathbf{H}$.

\section{Experiments}
\label{sec:ex}

\subsection{Datasets}
Since the raw $k$-space data can better verify the validity of the experiment than the amplitude image, we use three $k$-space raw datasets to evaluate the effectiveness of our approach.

\textbf{fastMRI:} This dataset is currently the largest open-access MRI dataset~\cite{zbontar2018fastmri}. Because there are no multi-modal data description on the official fastMRI website~\cite{xuan2020learning}, we select out $227$ pairs samples from the \texttt{training} set and $24$ samples pairs for \texttt{validation} set, including PD and FS-PDWI volumes which have been aligned. Since this dataset does not release the test set, we just report the results on the \texttt{validation} set.

\textbf{SMS:} We collected this dataset from 155 patients using a 3T Siemens Magnetom Skyra system. Each sample contains a T1WI and a T2WI with full $k$-space sampling, resulting in 2D volumes of size 320$\times$320 with 20 slices for each patient. The detailed acquisition parameters are as follows: TR$_{T1}$ = 2001ms, slice thickness = 5 mm, field of view (FOV) = 230$\times$200 mm$^2$, TE$_{T1}$ = 10.72 ms, TE$_{T2}$ = 112.86 ms, TR$_{T2}$ = 4511 ms.

\textbf{uMR:} We collected this dataset from 50 patients with a 3T whole body scanner uMR 790
\footnote{Provided by {\href{United Imaging Healthcare, Shanghai, China}{United Imaging Healthcare, Shanghai, China}.}}. Each sample contains a T1WI and a T2WI with full $k$-space sampling, resulting in 2D volumes of size 320$\times$320 with 19 slices for each patient. The detailed acquisition parameters are as follows: TR$_{T1}$ = 2001 ms, FOV = 220$\times$250 mm$^2$, slice thickness = 4 mm, matrix = 320$\times$320$\times$19, TR$_{T1}$ = 4511 ms, TE$_{T1}$ = 10.72 ms, TE$_{T2}$ = 112.86 ms.

Each dataset is aligned through the simple affine registration algorithm\footnote{{\href{https://github.com/Slicer/Slicer-RigidRegistration}{https://github.com/Slicer/Slicer-RigidRegistration}.}} before our experiments. SMS and uMRI are split patient-wise with a ratio of 7:1:2 for the \texttt{train/validation/test} sets. Finally, we obtain the LR image by truncating the outer part of the $k$-space~\cite{chen2018efficient} according to the enlargement scale factor, see \S\ref{arc} for more details.

\subsection{Baselines and Training Details}
We use PSNR and SSIM to evaluate the performance of SANet. The results are compared with the following previous works, including single-contrast and multi-contrast SR methods. The hyperparameter optimization for each method is performed by cross-validation. Single-contrast SR including EDSR~\cite{lim2017enhanced}, a classical SR method, the learning rate is initialized as $10^{-4}$ and Adam is adopted as the optimizer; SMORE~\cite{zhao2018deep}, an anti-aliasing method for MR image SR. Multi-contrast SR including Zeng \textit{et al.}~\cite{zeng2018simultaneous}, a CNN-based method which uses the T1WI as the prior information to guide the restoration of the target contrast, where the learning rate is initialized as $10^{-3}$ and Adam is adopted as the optimizer; Lyu \textit{et al.}~\cite{lyu2020multi}, a progressive network combine the auxiliary contrast in the high-level feature space for the restoration of the target contrast, where the learning rate is initialized as $10^{-5}$, the hyperparameter $\lambda_1$, $\lambda_2$, and $\lambda_3$ are set as 1.0, 0.1 and 0.01, respectively, and Adam is adopted as the optimizer; McMRSR~\cite{li2022transformer}, a multi-scale contextual matching and aggregation network for multi-contrast MR image super-resolution, where the learning rate is initialized as $10^{-4}$ and Adam is adopted as the optimizer; MINet~\cite{feng2021MINet}, our conference version, which trained with Adam optimizer and learning rate of $10^{-5}$, the parameters $\alpha$ and $L$ are empirically set to 0.7 and 6, respectively. Our model is trained using PyTorch with two NVIDIA Tesla V100 GPUs and 32GB of memory per card. We use the Adam optimizer with a learning rate of $10^{-5}$ for 50 epochs to train the model. The parameter $\alpha$ is empirically set to 0.7, the number of residual groups $L$ are set to 6. 
\begin{table*}[!t]
 \centering
 	\makeatletter\def\@captype{table}\makeatother\caption{Quantitative comparison results with standard deviation on three datasets with 2$\times$ enlargement. $P<$ 0.001 was considered as a statistically significant level.} 
	\resizebox{1\textwidth}{!}{
		\setlength\tabcolsep{3pt}
				\renewcommand\arraystretch{1.1}
				\begin{tabular}{r||ccc||ccc||ccc}
			        \hline\thickhline
			        \rowcolor{mygray}
			        {{}}&  \multicolumn{3}{c}{\textbf{fastMRI}} &\multicolumn{3}{c} {\textbf{SMS}} 
			        &\multicolumn{3}{c} {\textbf{uMRI}}\\\hline
			        \rowcolor{mygray}

			     {Method}
			        &~~PSNR$\uparrow$~~ &~~SSIM$\uparrow$~~ ~~&~~$P$ values~~&~~PSNR$\uparrow$~~ &~~SSIM$\uparrow$~~&~~$P$ values~~&~~PSNR$\uparrow$~~ &~~SSIM$\uparrow$~~&~~$P$ values~~  \\ \hline\hline
                    EDSR~\cite{lim2017enhanced}  &26.7$\pm$1.11 &0.512$\pm$0.06&$<0.001$/$<0.001$ &36.4$\pm$1.04 &0.962$\pm$0.06&$<0.001$/$<0.001$ &35.4$\pm$1.55 &0.965$\pm$0.07&$<0.001$/$<0.001$ \\
                    SMORE~\cite{zhao2018deep}  &28.3$\pm$1.00 &0.667$\pm$0.07&$<0.001$/$<0.001$ &38.1$\pm$1.08 &0.973$\pm$0.05&$<0.001$/$<0.001$ &36.4$\pm$1.30 &0.971$\pm$0.05&$<0.001$/$<0.001$\\
                    Zeng \textit{et al.}~\cite{zeng2018simultaneous} &28.9$\pm$1.05 &0.670$\pm$0.05 &$<0.001$/$<0.001$&38.2$\pm$1.10 &0.973$\pm$0.05 &$<0.001$/$<0.001$&36.4$\pm$1.20 &0.971$\pm$0.03&$<0.001$/$<0.001$\\
                    Lyu \textit{et al.}~\cite{lyu2020multi}  &29.5$\pm$0.80 &0.682$\pm$0.04 &$<0.001$/$<0.001$&39.2$\pm$0.94 &0.978$\pm$0.05 &$<0.001$/$<0.001$&37.1$\pm$1.09 &0.977$\pm$0.03&$<0.001$/$<0.001$\\
                    McMRSR~\cite{li2022transformer} &31.5$\pm$1.33 &0.699$\pm$0.04 &$<0.001$/$<0.001$&40.0$\pm$1.20 &0.963$\pm$0.06 &$<0.001$/$<0.001$&37.6$\pm$1.00 &0.946$\pm$0.03&$<0.001$/$<0.001$\\\hline\hline
                    MINet~\cite{feng2021MINet} &{\color{blue}31.8$\pm$0.83} &{\color{blue}0.709$\pm$0.03}&$<0.001$/$<0.001$ &{\color{blue}40.5$\pm$0.69} &{\color{blue}0.983$\pm$0.04}&$<0.001$/$<0.001$ &{\color{blue}38.0$\pm$1.00} &{\color{blue}0.980$\pm$0.03}&$<0.001$/$<0.001$\\ 
                    \textbf{SANet} &{\color{red}32.0$\pm$0.71} &{\color{red}0.716$\pm$0.03}&$<0.001$/$<0.001$ &{\color{red}42.1$\pm$0.59} &{\color{red}0.997$\pm$0.03}&$<0.001$/$<0.001$ &{\color{red}39.9$\pm$0.90} &{\color{red}0.990$\pm$0.03}&$<0.001$/$<0.001$\\ \hline\hline

 \end{tabular}}
 \label{table1}
\end{table*}

\begin{table*}[!t]
 \centering
 	\makeatletter\def\@captype{table}\makeatother\caption{Quantitative comparison results with standard deviation on three datasets with 4$\times$ enlargement. $P<$ 0.001 was considered as a statistically significant level.}  
	\resizebox{1\textwidth}{!}{
		\setlength\tabcolsep{3pt}
				\renewcommand\arraystretch{1.1}
				\begin{tabular}{r||ccc||ccc||ccc}
			        \hline\thickhline
			        \rowcolor{mygray}
			        {{}}&  \multicolumn{3}{c}{\textbf{fastMRI}} &\multicolumn{3}{c} {\textbf{SMS}} 
			        &\multicolumn{3}{c} {\textbf{uMRI}}\\\hline
			        \rowcolor{mygray}

			     {Method}
			        &~~PSNR$\uparrow$~~ &~~SSIM$\uparrow$~~ ~~&~~$P$ values~~&~~PSNR$\uparrow$~~ &~~SSIM$\uparrow$~~&~~$P$ values~~&~~PSNR$\uparrow$~~ &~~SSIM$\uparrow$~~&~~$P$ values~~  \\ \hline\hline
                    EDSR~\cite{lim2017enhanced}  &18.4$\pm$1.02 &0.208$\pm$0.06&$<0.001$/$<0.001$ &31.5$\pm$1.19 &0.886$\pm$0.07&$<0.001$/$<0.001$ &31.2$\pm$1.29 &0.907$\pm$0.07&$<0.001$/$<0.001$\\
                    SMORE~\cite{zhao2018deep}  &21.8$\pm$0.95 &0.476$\pm$0.07&$<0.001$/$<0.001$ &32.1$\pm$1.00 &0.901$\pm$0.06&$<0.001$/$<0.001$ &32.0$\pm$1.17 &0.918$\pm$0.05&$<0.001$/$<0.001$\\
                    Zeng \textit{et al.}~\cite{zeng2018simultaneous}  &23.3$\pm$0.91 &0.507$\pm$0.06 &$<0.001$/$<0.001$&32.5$\pm$1.03 &0.912$\pm$0.04 &$<0.001$/$<0.001$&31.9$\pm$1.22 &0.921$\pm$0.06&$<0.001$/$<0.001$\\
                    Lyu \textit{et al.}~\cite{lyu2020multi}  &28.2$\pm$0.86 &0.574$\pm$0.06 &$<0.001$/$<0.001$&33.7$\pm$1.00 &0.931$\pm$0.04 &$<0.001$/$<0.001$&32.2$\pm$1.02 &0.929$\pm$0.03&$<0.001$/$<0.001$\\
                    McMRSR~\cite{li2022transformer} &{\color{blue}29.9$\pm$1.05} &{\color{blue}0.609$\pm$0.05} &$<0.001$/$<0.001$&34.6$\pm$0.89 &0.945$\pm$0.04 &$<0.001$/$<0.001$&34.0$\pm$1.12 &0.951$\pm$0.03&$<0.001$/$<0.001$\\\hline\hline
                    MINet~\cite{feng2021MINet} &29.8$\pm$0.90 &0.601$\pm$0.04&$<0.001$/$<0.001$ &{\color{blue}35.0$\pm$0.92} &{\color{blue}0.948$\pm$0.03}&$<0.001$/$<0.001$ &{\color{blue}34.2$\pm$1.01} &{\color{blue}0.956$\pm$0.03}&$<0.001$/$<0.001$\\ 
                    \textbf{SANet} &{\color{red}30.4$\pm$0.83} &{\color{red}0.622$\pm$0.04}&$<0.001$/$<0.001$ &{\color{red}35.6$\pm$0.70} &{\color{red}0.957$\pm$0.03}&$<0.001$/$<0.001$ &{\color{red}35.3$\pm$0.79} &{\color{red}0.963$\pm$0.03}&$<0.001$/$<0.001$\\ \hline\hline
 \end{tabular}}
 \label{table2}
\end{table*}

\begin{figure*}[!t]
\centering
  \includegraphics[width=\linewidth]{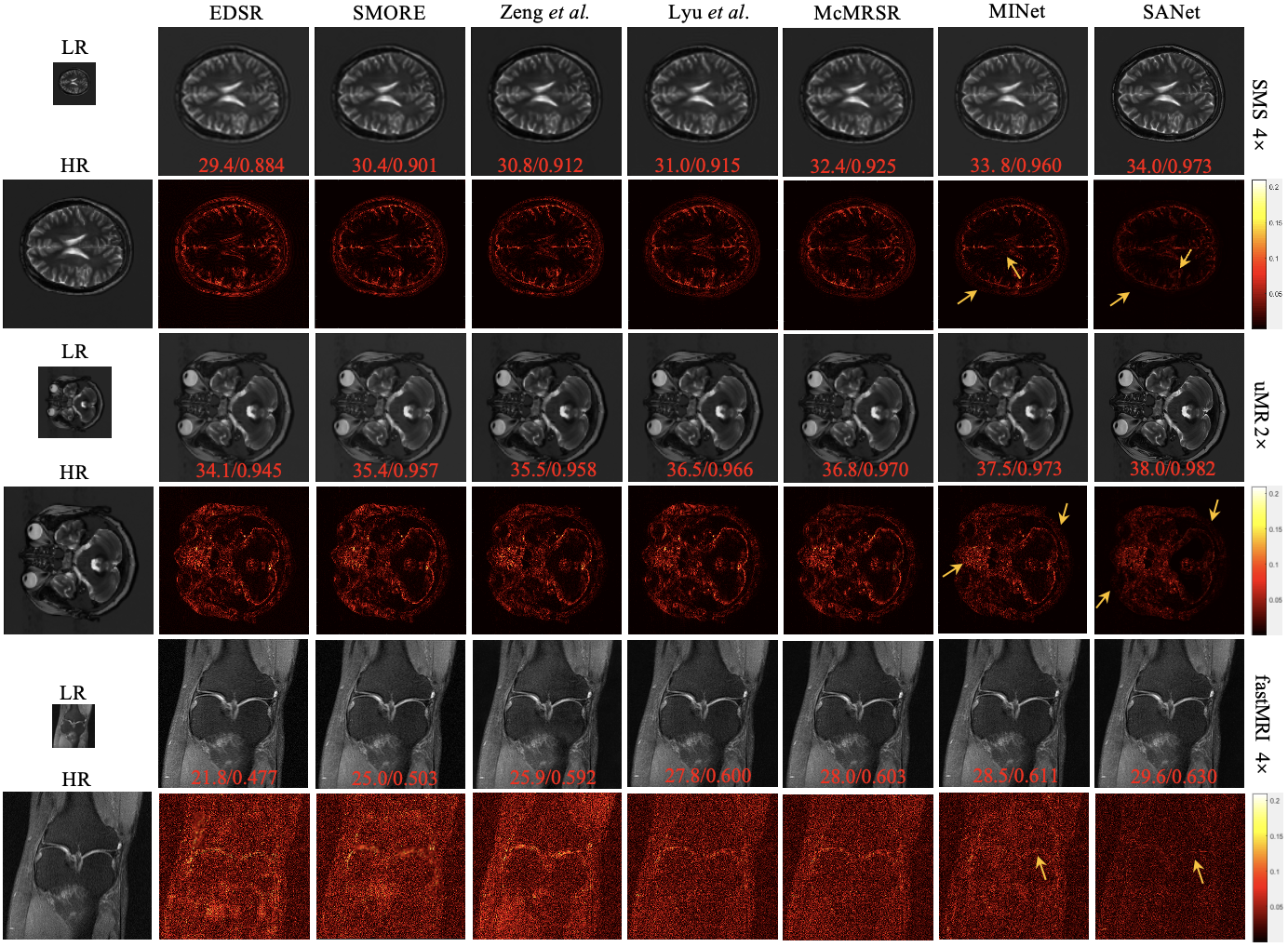}
  \caption{Qualitative comparison of various methods on the three datasets, with \textbf{2$\times$} and \textbf{4$\times$} enlargement. SR images and error maps are presented with their PSNR and SSIM values.}
  \label{fig01}
\end{figure*}

\subsection{Quantitative Evaluation }

Here, we use the average SSIM and PSNR results calculated for the fully sampled images as the ground-truths. Table~\ref{table1} reports the target contrast restoration evaluations under 2$\times$ enlargement. Since the diverse machines used for each dataset, there are some discrepancies in their results. Notably, the best results come from our method both on the three datasets. It indicate that our method can effectively guide the restoration of the target contrast. However, since the lack of auxiliary contrast information in the single-mode methods, the results of EDSR~\cite{lim2017enhanced} and SMORE~\cite{zhao2018deep} are far less effective than other methods. similarly, Zeng \textit{et al.}~\cite{zeng2018simultaneous} and Lyu \textit{et al.}~\cite{lyu2020multi}, do not employ a deep fusion mechanism, \eg, the dependencies between different contrast at each stage and the high/low-intensity attention, they are less effective than our model. Although McMRSR~\cite{li2022transformer} adopts a multi-scale context matching and aggregation scheme, it still has some shortcomings in capturing the structure edge details. Notably, MINet is also obviously superior to any comparison method. Nevertheless, with the help of the separable attention, our results are further enhanced. For example, we improve the PSNR from 29.5 dB to 31.8 dB with MINet and to 32.0 dB with SANet, and the SSIM from 0.682 to 0.709 with MINet and to 0.716 with SANet on the fastMRI dataset, as compared to the previous best method, \ie, Lyu \textit{et al.}~\cite{lyu2020multi}. 

Table~\ref{table2} reports the target contrast restoration evaluations under 4$\times$ enlargement. As can be seen, compared to the PSNR of SMORE~\cite{zhao2018deep} (21.8/32.1/32.0 dB), which dose not use the auxiliary contrast, our MINet and SANet achieve significantly better results. Compared to the previous state-of-the-art methods that do use the auxiliary contrast,~\ie, Zeng \textit{et al.}~\cite{zeng2018simultaneous} and Lyu \textit{et al.}~\cite{lyu2020multi}, we improve the results from 28.2/33.7/32.2 dB to 30.4/35.6/35.3 dB. Our multi-contrast method with its multi-stage integration module is thus clearly effective for target-contrast image restoration, while the separable attention greatly improves the final results. Because of the powerful multi-contrast fusion capability of our method, it is superior to previous methods in both 2$\times$ and 4$\times$ enlargement scenarios. To substantiate our results, we used a paired Student’s t-test to evaluate the significant difference between the two methods~\cite{ye2019deep}. $P<$ 0.001 was considered as a statistically significant level. As can be seen from the $P$ values in Tables~\ref{table1} and Table~\ref{table2}, our method has a statistically significant improvement over the various baselines.

\subsection{Qualitative Evaluation } \label{sec:error}
The qualitative comparison results on the different enlarge scales of SR images and their corresponding error maps for the three datasets are shown~\figref{fig01}. The texture on an error map represents the restoration error; the smoother the texture, the better the restoration. As shown in this figure, the error of multi-contrast methods is lower than that of single-contrast models. With the increase in magnification, the difficulty of image restoration also increases. It is worth noting that the images recovered by our MINet and SANet have less of a checkerboard effect and less structural loss. In particular, SANet significantly \textit{reduces the error on the key organizational and structural margins} (see the yellow arrows in~\figref{fig01}), thanks to the separable attention. 
\begin{figure*}[!htb]
\centering
  \includegraphics[width=0.85\textwidth]{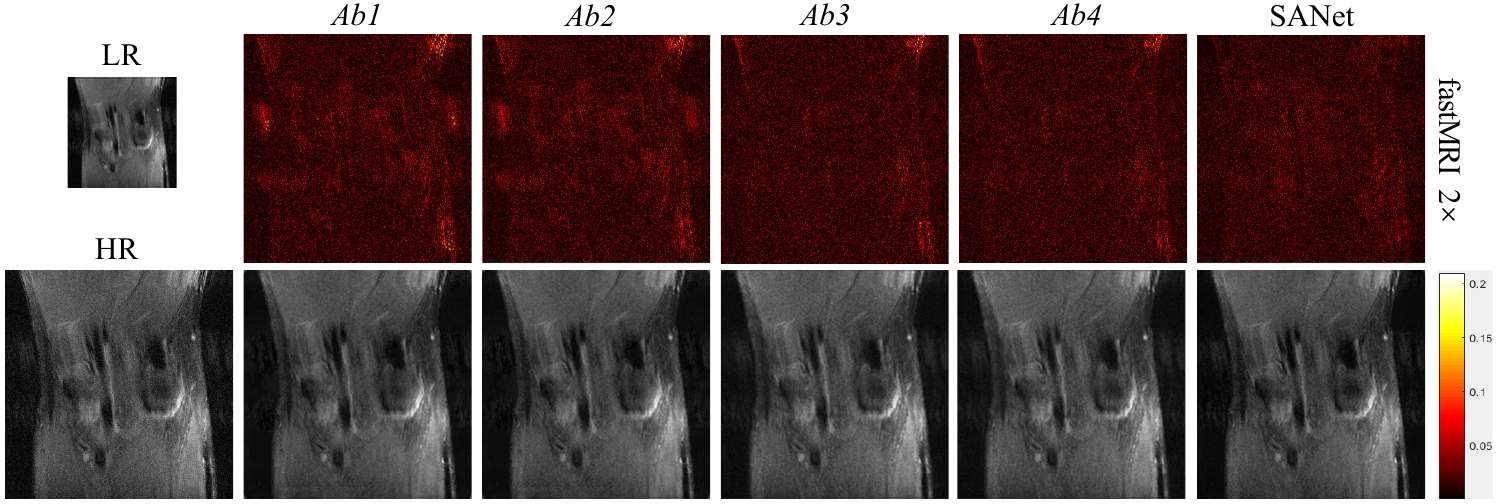}
  \caption{Ablation study of the key parts in our method, where $Ab_1$, $Ab_2$, $Ab_3$, and $Ab_4$ are the derived models with different key components.}
  \label{fig02}
\end{figure*}

\subsection{Ablation Studies } \label{sec:ablation}

We discuss our method through various aspects, including an analysis of the key components, a discussion of the auxiliary modality, a comparison of the number of blocks, parameter analysis, and the analysis of multi-contrast for joint super-resolution.

\begin{table*}[!t]   
 \centering
 \caption{Ablation study regarding the key components of our SANet on the {SMS} dataset with 2$\times$ enlargement. $P<$ 0.001 was considered as a statistically significant level.}
 \resizebox{0.95\linewidth}{!}{
 \setlength\tabcolsep{4pt}
 \renewcommand\arraystretch{1.0}
				\begin{tabular}{r||cccc|cccc}
			        \hline
			        \hline
			        \thickhline
			        \rowcolor{mygray}

			        {Variant} &~~$\mathcal{F}_{\text{Aux}}$~~ &~~$\mathcal{M}_{\text{Int}}$~~ & ~~$\mathcal{M}_{\text{Att}}$~~ &~~$\mathcal{S}^l$~~&~~PSNR~~&~~SSIM~~&~~NMSE~~ ~~&~~ $P$ valus\\ \hline\hline
                    $Ab_1$ &\XSolidBrush &\XSolidBrush & \XSolidBrush &\XSolidBrush &38.6$\pm$0.015 &0.970$\pm$0.009 &0.004$\pm$0.012&$<0.001$/$<0.001$/$<0.001$ \\ 
                    $Ab_2$ &\XSolidBrush &\XSolidBrush &\Checkmark &\XSolidBrush &39.2$\pm$0.108 &0.978$\pm$0.025 &0.003$\pm$0.002&$<0.001$/$<0.001$/$<0.001$ \\ 
                    $Ab_3$ &\Checkmark &\XSolidBrush & \Checkmark &\XSolidBrush &39.2$\pm$0.001 &0.979$\pm$0.037 &0.002$\pm$0.036&$<0.001$/$<0.001$/$<0.001$\\
                    $Ab_4$  &\Checkmark &\Checkmark & \Checkmark &\XSolidBrush &{\color{blue}40.5$\pm$0.014} &{\color{blue}0.983$\pm$0.009} &{\color{blue}0.002$\pm$0.004} &$<0.001$/$<0.001$/$<0.001$\\ 
                    {SANet}&\Checkmark &\Checkmark & \Checkmark &\Checkmark &{\color{red}41.0$\pm$0.038} &{\color{red}0.990$\pm$0.010} &{\color{red}0.002$\pm$0.002} &$<0.001$/$<0.001$/$<0.001$  \\ \hline\hline
	        	 \end{tabular}}
 \label{ab}	        	 
\end{table*}

\vspace*{3pt} 
\subsubsection{Analysis of the key components.}
In this section, we first evaluate the performance of the key components of our model. The performance of four important components is evaluated under 2$\times$ enlargement. The components evaluated include $\mathcal{F}_{\text{Aux}}$, which is the auxiliary contrast in our model, $\mathcal{M}_{\text{Int}}$, which is our multi-stage integration module, $\mathcal{M}_{\text{Att}}$, which is our multi-contrast feature enhancement module, and $\mathcal{S}^l$, which is our separable attention compensation module. Table~\ref{ab} summarizes the component analysis under 2$\times$ enlargement on the $\textbf{SMS}$ dataset, where $Ab_1$, $Ab_2$, $Ab_3$, and $Ab_4$ are the four derived ablation models. From this table, we observe that $Ab_2$ improves the performance over the single-contrast version, $Ab_1$, by 0.6 dB, demonstrating the importance of the auxiliary contrast. Combining the auxiliary contrast, $\mathcal{M}_{\text{Int}}$ and $\mathcal{M}_{\text{Att}}$ achieve the largest boost in performance, \eg, 38.6 dB to 40.5 dB, as compared to the baseline. This supports our conclusion that these two modules can provide deep information to complement the SR of the target contrast. Our SANet, equipped with all components, produces the best restoration results, improving the PSNR from 38.6 dB to 41.0 dB. Overall, all four components help enhance the performance of SANet, providing it with a powerful capability of extract important information to guide the SR of target-contrast. It should be noted that although $\mathcal{S}^l$ can further improve the reconstruction results than other fusion scheme, $\mathcal{S}^l$ is also built on the auxiliary contrast, demonstrating  that our separable attention mechanism is an effective multi-contrast assisted imaging scheme.

\subsubsection{Analysis of the Separable Attention Compensation Module}
\begin{figure}[!t]
\centering
  \includegraphics[width=0.5\textwidth]{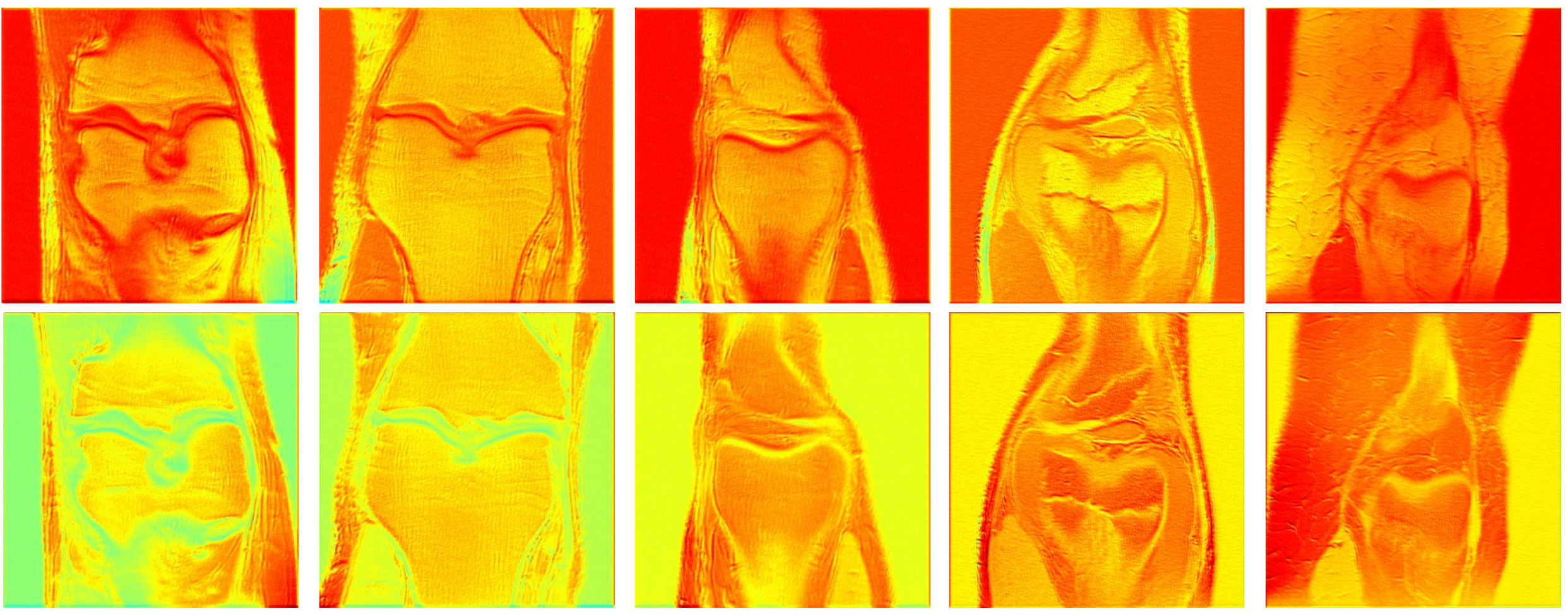}
  \caption{Separable attention visualization, where the first line represents high-intensity priority (HP) attention map, and the second line represents low-intensity separation (LS) attention map. The brighter the color, the higher the network's attention.}
  \label{attmap} 
\end{figure}

Here, we will focus on the analysis of why our separable attention compensation module is useful for MR image restoration. We have mentioned that it can explore high-intensity and low-intensity areas in both forward and reverse directions, making the network pay more attention to the uncertain boundary details as well as anatomical details. We have shown the high-intensity priority and low-intensity separation feature maps in~\figref{fig22}. It is obvious that the two feature maps are focused on the anatomical structure and edge details, respectively (see the {\color{cyan}blue} box). To better understand the separable attention compensation module, we show the high-intensity priority (HP) attention map and the low-intensity separation (LS) attention map of fastMRI dataset in~\figref{attmap}. The brighter the color, the higher the network's attention. As can be seen from this figure, HP focuses on the high-intensity structure, while LS focuses on the low-intensity region. In other words, such reverse attention mechanism can joint inferring to produce bidirectional enhanced visual features. More directly, compared the error map of MINet with SANet, we observed that SANet can recover the boundary and organizational details of the image, such as the yellow arrow area, with the help of the separable attention compensation module. 

\vspace*{3pt} 
\subsubsection{Discussion of the auxiliary modality.} 
\begin{figure}[t]
\centering
  \includegraphics[width=0.5\textwidth]{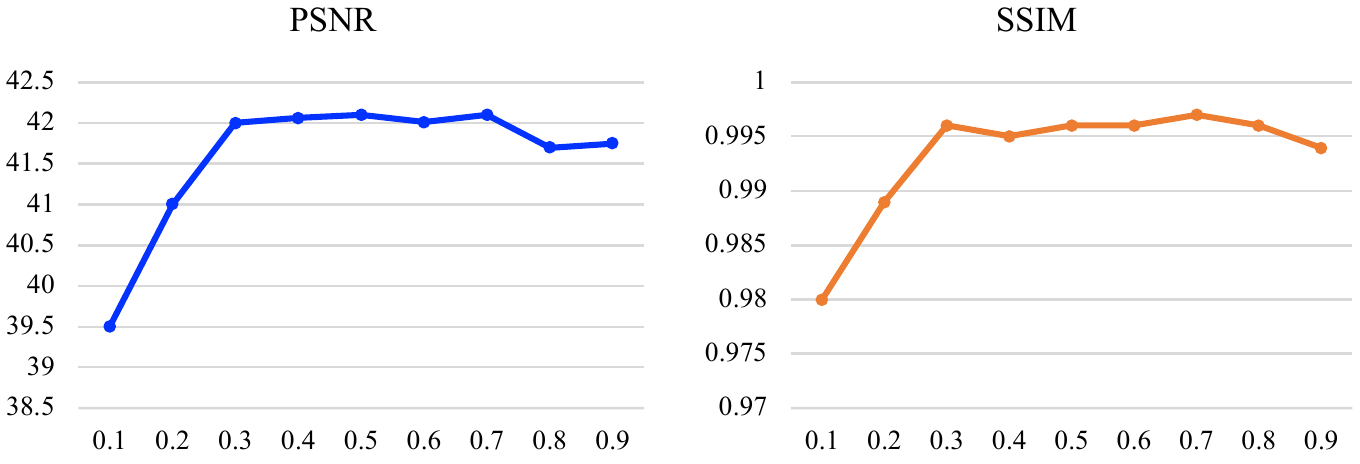}
  \caption{ Ablation study on the ratio of the two contrasts $\alpha$ in terms of PSNR and SSIM. The abscissa value is $\alpha$.}
  \label{figalpha}
\end{figure}

To investigate the influence of weighting the trade-off between the two modalities, we report the SR results of our method on the $\textbf{SMS}$ dataset with 2$\times$ enlargement in~\figref{figalpha}. The weights of both the target and auxiliary streams are determined by the value of $\alpha$. The bigger the value of $\alpha$, the smaller the influence of the auxiliary modality in our model. From this figure, we can see that our model has the best PSNR and SSIM results at $\alpha$ = 0.7, which demonstrates that the auxiliary contrast plays an important role in our model. However, when $\alpha$ is smaller than 0.3, the PSNR and SSIM values quickly degrades. This is likely because too small a value of $\alpha$, results in insufficient target contrast information.

\begin{figure}[t]
\centering
  \includegraphics[width=0.5\textwidth]{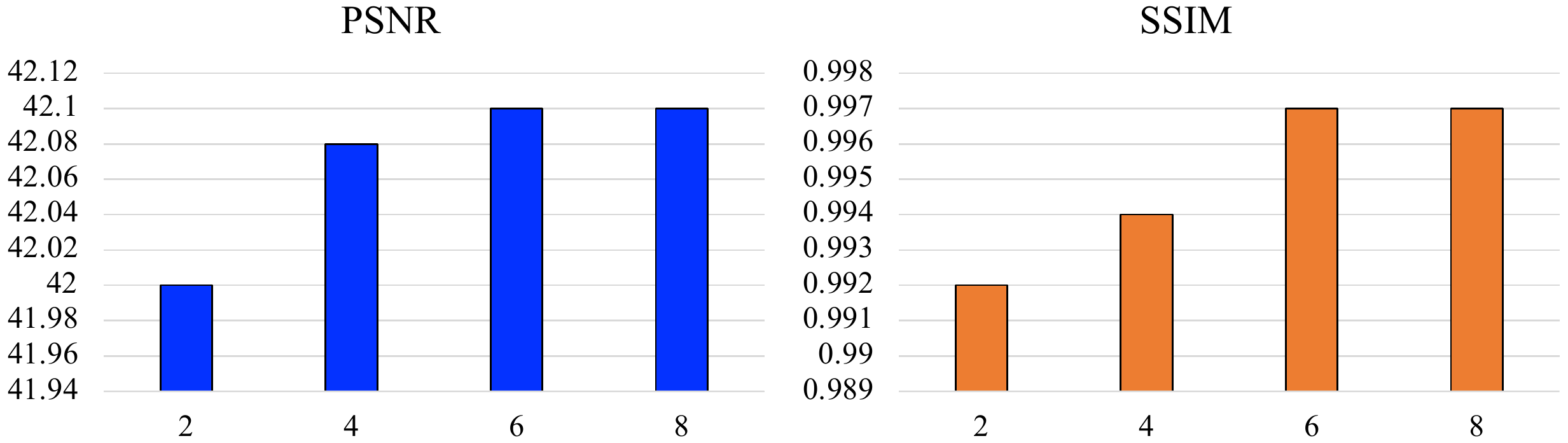}
  \caption{Ablation study on using different numbers of residual groups. The abscissa value is the number of residual groups.}
  \label{figgroup}
\end{figure}

\vspace*{3pt} 
\subsubsection{Comparison of the number of blocks.} 
As the number of network parameters grows with the number of residual groups, we need to select an appropriate number of blocks to balance the accuracy and computational complexity of the network. In~\figref{figgroup}, we evaluate the PSNR and SSIM scores of different numbers groups on the $\textbf{SMS}$ dataset with 2$\times$ enlargement. As shown in this figure, the SR performance, measured by PSNR and SSIM, increases monotonically when the number of residual groups increases, becoming stable when the number of groups is 6. However, although it uses fewer numbers of groups, our method still generates higher PSNR and SSIM scores than other baselines (see Table~\ref{table1}). This study reveals the effectiveness of our proposed separable attention and multi-stage integration modules.

\vspace*{3pt} 
\subsubsection{Parameter analysis.} 
We also find that our method requires the lowest number of parameters. For example, SANet requires 5M parameters when residual groups = 2, and 11M when residual groups = 6, where the separable attention module requires 0.2M parameters and the FLOPs of this module are 17.77G. The FLOPs of SANet are 307.02G and the execution time is 2.20s/it. However, EDSR requires 43M parameters, and both SMORE~\cite{zhao2018deep} and Zeng \textit{et al.}~\cite{zeng2018simultaneous} were built on the framework of EDSR, among which SMORE is composed of two EDSR framework. Lyu \textit{et al.}~\cite{lyu2020multi} adopted progressive model which requires 7M parameters, but the performance worse than our method.

\vspace*{3pt} 
\subsubsection{Analysis of Multi-Contrast for Joint Super-Resolution.}
To reveal the effectiveness of our SANet for multi-contrast fusion, we record the values where the two contrasts need to be super-resolved simultaneously. The results of SANet are PSNR = 31.7.2 dB and SSIM = 0.700 in the target contrast, PSNR = 32.3 dB and SSIM = 0.705 in the auxiliary contrast on the fastMRI data with 2$\times$ enlargement. These results are significantly higher than the baseline, as shown in Table~\ref{table1}. Thus, SANet is a powerful multi-contrast fusion mechanism, which can be used to restore images of multiple contrasts simultaneously.

\subsubsection{Discussion of the Loss Function.}

In Table~\ref{table:4}, we record the 2$\times$ enlargement results of two commonly used loss on the SMS dataset, \eg, $L_1$ and $L_2$. The results in this table show that there is no significant difference between the restoration results under the two losses~\cite{ledig2017photo}. Since the convergence speed of $L_1$ is faster than $L_2$ loss~\cite{lim2017enhanced}, the $L_1$ loss is adopted in our model.

\begin{table}[!t]
	\centering
	\caption{Discussion of the loss function on {\textbf{SMS}} datasets with 2$\times$ enlargement.}
	\resizebox{0.5\textwidth}{!}{
		\setlength\tabcolsep{10pt}
		\renewcommand\arraystretch{1.0}
		\begin{tabular}{r||ccc}
			\hline\thickhline
			\rowcolor{mygray}
			Loss & PSNR &SSIM &$P$ values \\ \hline\hline
			$L_1$ &42.10$\pm$0.59 &0.997$\pm$0.03 &$<0.001$/$<0.001$   \\
			$L_2$ &42.13$\pm$0.62 &0.998$\pm$0.05 &$<0.001$/$<0.001$   \\
			\hline
		\end{tabular}
	}
	\captionsetup{font=small}

	\label{table:4} 
\end{table}

\section{Conclusion}
In this work, we aim to explore the high-intensity and low-intensity relationships in the auxiliary contrast to assist the super-resolution of the target contrast. For this purpose, we propose a separable attention mechanism, named SANet, for multi-contrast MR image super-resolution, which can effectively restore the target contrast under the guidance of the auxiliary one. In this method, a high-intensity priority attention and a low-intensity separation attention are used to jointly reconstruct the image from the complementary information of the high-intensity and low-intensity regions in the auxiliary contrast. In this way, more attention is diverted to refining the uncertain details between the two regions, yielding a higher-quality target-contrast image. In addition, we design a multi-stage integration module to explore the responses of the multi-contrast fusion at different stages and improve their representation ability. We conduct numerous experiments on various datasets, demonstrating the superiority of our model over the most advanced methods for multi-contrast MR image super-resolution. 


\bibliographystyle{IEEEtran}
\bibliography{ref.bib}

\vspace{-8mm}
\begin{IEEEbiography}[{\includegraphics[width=1in,height=1.25in,clip,keepaspectratio]{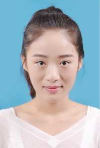}}]{Chun-Mei Feng}
	\small received her M.S. degree at QuFu Normal University, China, in 2018. She is currently a Ph.D. student of the school of Information Science and Technology, Harbin Institute of Technology Shenzhen, China. Her research interests include medical imaging and bioinformatics.
\end{IEEEbiography}
\vspace{-8mm}
\begin{IEEEbiography}[{\includegraphics[width=1in,height=1.25in,clip,keepaspectratio]{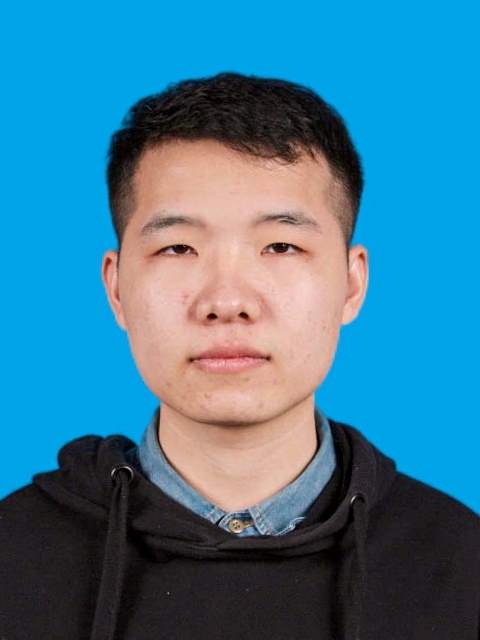}}]{Yunlu Yan}
	\small received his Bachelor's degree at NanChang University, China, in 2020. He is currently a M.S. student of the school of Information Science and Technology, Harbin Institute of Technology Shenzhen, China. His research interests include medical imaging and federated learning.
\end{IEEEbiography}
\vspace{-8mm}
\begin{IEEEbiography}[{\includegraphics[width=1in,height=1.25in,clip,keepaspectratio]{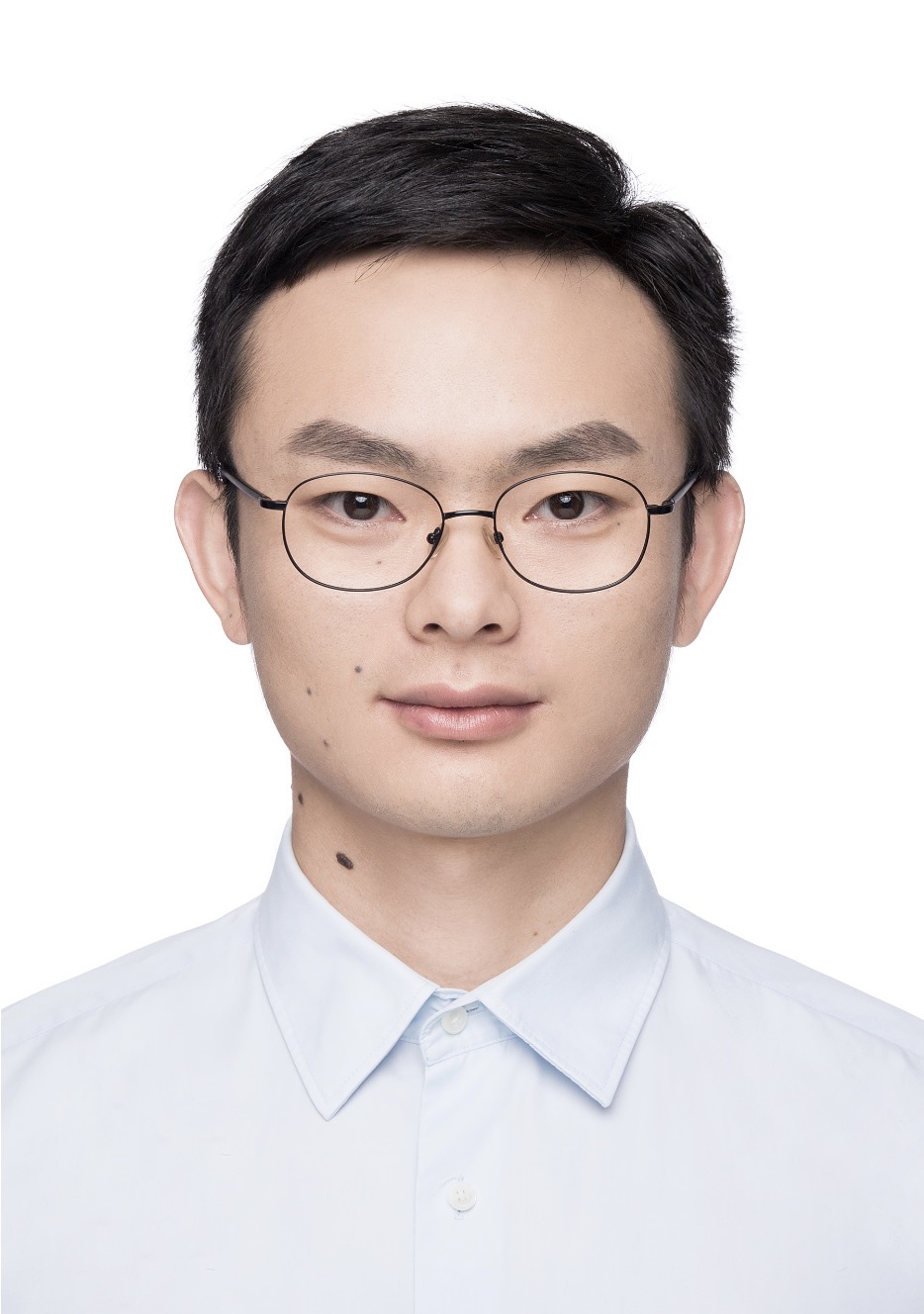}}]{Kai Yu}
	\small is a Scientist at the Institute of High Performance Computing, A*STAR, Singapore. He received his Ph.D. from Soochow University in 2020 and was an assistant researcher at the Children's Hospital of Zhejiang University School of Medicine for two years. His research interests include computer vision, medical image analysis, and heart sound signal analysis.
\end{IEEEbiography}
\vspace{-8mm}
\begin{IEEEbiography}[{\includegraphics[width=1in,height=1.25in,clip,keepaspectratio]{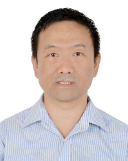}}]{Yong Xu}
	\small (Senior Member, IEEE) received his B.S. and M.S. degrees at Air Force Institute of Meteorology (China) in 1994 and1997, respectively. He then received his Ph.D. degree in pattern recognition and intelligence system at the Nanjing University of Science and Technology in 2005. From May 2005 to April 2007, he worked at Harbin Institute of Technology Shenzhen as a post-doctoral research fellow. Now he is a professor at HIT Shenzhen. His current interests include pattern recognition, machine learning, and bioinformatics.
\end{IEEEbiography}
\vspace{-8mm}
\begin{IEEEbiography}[{\includegraphics[width=1in,height=1.25in,clip,keepaspectratio]{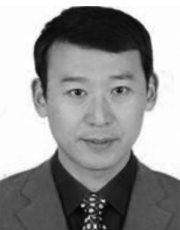}}]{Jian Yang}
	\small (Member, IEEE) received the PhD degree from Nanjing University of Science and Technology (NUST), on the subject of pattern recognition and intelligence systems in 2002. In 2003, he was a Postdoctoral researcher at the University of Zaragoza. From 2004 to 2006, he was a Postdoctoral Fellow at Biometrics Centre of Hong Kong Polytechnic University. From 2006 to 2007, he was a Postdoctoral Fellow at Department of Computer Science of New Jersey Institute of Technology. Now, he is a Chang-Jiang professor in the School of Computer Science and Technology of NUST. He is the author of more than 200 scientific papers in pattern recognition and computer vision. His papers have been cited more than 6000 times in the Web of Science, and 18000 times in the Scholar Google. His research interests include pattern recognition, computer vision and machine learning. Currently, he is/was an associate editor of Pattern Recognition, Pattern Recognition Letters, IEEE Trans. Neural Networks and Learning Systems, and Neurocomputing. He is a Fellow of IAPR.
\end{IEEEbiography}
\vspace{-8mm}
\begin{IEEEbiography}[{\includegraphics[width=1in,height=1.25in,clip,keepaspectratio]{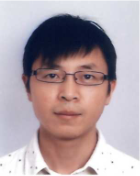}}]{Ling Shao}
	\small (Fellow, IEEE) is the Chief Scientist and President of Terminus International, China. He is an Associate Editor of the IEEE TRANSACTIONS ON IMAGE PROCESSING, the IEEE TRANSACTIONS ON NEURAL NETWORKS AND LEARNING SYSTEMS, the IEEE TRANSACTIONS ON CIRCUITS ANDSYSTEM FOR VIDEO TECHNOLOGY, and several other journals. His research interests include Computer Vision, Machine Learning and Medical Imaging. He is a Fellow of the IEEE, the IAPR, the IET and the BCS.
\end{IEEEbiography}
\vspace{-8mm}
\begin{IEEEbiography}[{\includegraphics[width=1in,height=1.25in,clip,keepaspectratio]{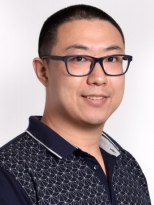}}]{Huazhu Fu}
	\small (SM'18) is a Senior Scientist at the Institute of High Performance Computing, A*STAR, Singapore. He received his Ph.D. from Tianjin University in 2013, and was a Research Fellow at Nanyang Technological University for two years. From 2015 to 2018, he was a Research Scientist in Institute for Infocomm Research at Agency for Science, Technology and Research. His research interests include computer vision, machine learning, and medical image analysis. He currently serves as an Associate Editor of IEEE Transactions on Medical Imaging, IEEE Journal of Biomedical and Health Informatics, and IEEE Access. He also serves as a co-chair of OMIA workshop and co-organizer of ocular image series challenge (i-Challenge). 
\end{IEEEbiography}
\end{document}